%% file: main.tex
\documentclass[lettersize,journal]{IEEEtran}
\usepackage{cite}
\usepackage{amsmath,amssymb,amsfonts}
\usepackage{algorithmicx}
\usepackage{graphicx}
\usepackage{textcomp}
\usepackage[nolist]{acronym}
\usepackage{longtable}
\usepackage{relsize}
\usepackage{threeparttable}
\usepackage[normalem]{ulem}
\usepackage{comment}
\usepackage{color,soul}
\usepackage{multicol,multirow,xcolor,colortbl}
\usepackage{makecell}
\usepackage{orcidlink}
\usepackage{xparse}
\usepackage{caption}
\usepackage{subcaption}
\usepackage{algpseudocode}
\usepackage{svg}

\newcommand{\rev}[1]{\textcolor{black}{#1}}

\usepackage{xcolor}
\usepackage{etoolbox}

\input{acronyms}
\input{symbols}
\begin{document}

\title{Analysis of Proactive Uncoordinated Techniques to Mitigate Interference in FMCW Automotive Radars}

\author{
\IEEEauthorblockN{
Alessandro Bazzi\IEEEauthorrefmark{1}\IEEEauthorrefmark{2}, Francesco Miccoli\IEEEauthorrefmark{2},
Fabrizio Cuccoli\IEEEauthorrefmark{3}, Luca Facheris\IEEEauthorrefmark{3}\IEEEauthorrefmark{4}, and Vincent Martinez\IEEEauthorrefmark{5}
 }\\
\IEEEauthorblockA{%
\IEEEauthorrefmark{1}DEI, Universit\`a di Bologna, Italy\\
  \IEEEauthorrefmark{2}National Laboratory of Wireless Communications (WiLab), CNIT, Italy\\
  \IEEEauthorrefmark{3}National Laboratory of Radar and Surveillance Systems (RaSS), CNIT, Italy\\
  \IEEEauthorrefmark{4}DINFO, Universit\`a di Firenze, Italy\\
  \IEEEauthorrefmark{5}NXP Semiconductors, France\\
}
\thanks{Corresponding author: F. Miccoli (email: francesco.miccoli@wilab.cnit.it).}
\thanks{©2026 IEEE.  Personal use of this material is permitted.  Permission from IEEE must be obtained for all other uses, in any current or future media, including reprinting/republishing this material for advertising or promotional purposes, creating new collective works, for resale or redistribution to servers or lists, or reuse of any copyrighted component of this work in other works.}
}



\markboth{IEEE Transactions on Radar Systems, accepted for publication, 2026}%
{Shell \MakeLowercase{\textit{et al.}}: A Sample Article Using IEEEtran.cls for IEEE Journals}

\maketitle

\begin{abstract}
Modern vehicles increasingly rely on advanced driver-assistance systems (ADAS), with radars playing a key role due to their cost-effectiveness and reliable performance. However, the growing number of radars operating in the same spectrum raises concerns about mutual interference, which could lead to system malfunctions and potential safety risks. 
This study focuses on a scenario in which all vehicles are equipped with frequency-modulated continuous-wave (FMCW) radars, and it assesses the impact of interference on radar functionality — expressed in terms of probability of failure — by considering both direct and reflected signals. The radars may employ one of the following proactive mitigation methods to reduce the impact of interference, all of which require no inter-vehicle coordination but differ in complexity: (i) random carrier-frequency hopping on a frame-by-frame basis, (ii) random carrier-frequency hopping on a chirp-by-chirp basis, and (iii) a directional, compass-based method specifically addressing interference from opposite directions, which can be combined with either of the two previous methods. In this work, we assume realistic simulated road traffic scenarios and develop a novel model that captures correlated interference and accounts for the main radar setting parameters.
Results reveal that dense scenarios pose a high risk of radar malfunctions. Among the analyzed methods, chirp-by-chirp frequency hopping emerges as the most effective approach to mitigate interference and ensure system reliability, but only when combined with a sufficiently large bandwidth. The compass-based method, on the other hand, shows limited effectiveness and appears not worth the additional system complexity.
\end{abstract}


%
%

\section{Introduction}



To increase road safety, regulators are introducing mandates for an increasing number of \ac{ADAS} capabilities. This, in turn, necessitates the installation of an increasing number of sensors on board the vehicles, among which  \ac{FMCW} radars play a major role. The reason is that FMCW radars are relatively inexpensive and provide good accuracy and resolution for both distance and speed estimations, while also providing better resiliency to adverse weather conditions than LiDARs and cameras. 

An aspect that is not critical today but \rev{is considered by the automotive sector to} become problematic with their increased adoption is that radars currently share a limited spectrum, mostly between 77 and 81 GHz, with loose regulations and no coordination \rev{\cite{NHSTA_812_632}}. In the future, this may lead  to high levels of interference, potentially causing frequent malfunctioning of the systems. 
\rev{Indeed, an increasing number of papers and research project activities are addressing this issue and exploring ideas to mitigate the impact of interference, as reported, for example, in \rev{\cite{SAE_Meeting_Report,https://doi.org/10.1049/rsn2.12096,11241101,8828025}}. 
Additionally, for this reason the automotive sector is proposing the use of new frequency bands, with particular reference to a large portion of the spectrum around 140 GHz, as discussed in \cite{10734940,ECCreport351} and explored in ETSI TR 104 054.} 

Regarding the possible interference mitigation methods, they can be categorized into two main classes. The first class involves elaborating the signal a posteriori, also called \textit{reactive}. This approach, with examples provided in \cite{
9764338,8378800,8767459,
8448223}, can only attempt to reduce the impact of interference once it has already occurred and will not be further discussed in this paper. The second class, known as \textit{proactive} methods, acts on signals a priori. When proactive solutions are implemented, the reactive ones can still be applied to address any residual interference. 

When considering FMCW radar systems, another distinction is made between interference caused by radars featuring different or similar parameterizations, referred to as \textit{uncorrelated} and \textit{correlated} interference, respectively. Uncorrelated interference typically increases the noise floor, reducing the radar's detection range. In contrast, correlated interference can generate signals at the receiver that resemble reflections, potentially leading to the detection of false (ghost) objects. 
Even if different radar module vendors or car manufacturers may use different parameterizations, the number of configurations used in practice is limited, and correlated interference is likely to occur with high probability and may constitute a high safety risk.

To address the issue of interference, particularly when signals are correlated, several proactive strategies have been proposed in recent years for FMCW radars
\cite{7733011,9070137,9266425,8835744,8943325,roudiere2021first,roudiere2021importance,10044192,9023460,https://doi.org/10.1049/joe.2019.0166,9079146,
s18092831,6544299,8768092,8967012,9348730,9399786,9530166,5164283,10044244}. Some of these strategies have been developed within international projects such as MOSARIM and IMIKO \cite{9127843}. 
As detailed in the following, the attention is here restricted to the use of time and frequency variability on a frame-by-frame or chirp-by-chirp basis, 
because they do not require coordination between radars or agreements among stakeholders. It is relevant to remark that these methods, despite their conceptual simplicity, 
may pose practical implementation challenges; therefore, assessing their effectiveness in large-scale scenarios in advance is of crucial importance.

In this work, we consider a future scenario where all vehicles are equipped with \ac{FMCW} radars that may interfere with one another. We focus on realistic, dense highway scenarios reproduced with a road-traffic simulator, and assess the effectiveness of proactive mitigation methods to reduce the impact of correlated interference. 
\rev{Compared to the state of the art, this work introduces the following novelties:
\begin{itemize}
    \item To analyze the issue of interference independently of signal details, we propose the novel concept of \textit{potential interferers}, which identifies those radars that generate a signal that, directly or after reflection, enter in the \ac{FOV} of the radar under analysis; we use the term potential, because such signal may or may not actually interfere depending on the bandwidth and the other characteristics of the signal;
    \item We develop a model that statistically estimates the probability that the radar is blinded due to 
    correlated 
    interference, considering the realistic scenario, the available system bandwidth, and a large number of radar parameters;
    \item We compare different mitigation methods in highway scenarios with hundreds of vehicles, evaluating solutions that can be implemented
without altering radar performance during operation and without requiring cooperation among radars, namely (i) frame-by-frame random frequency hopping, (ii) chirp-by-chirp random frequency hopping, and (iii) a compass-based frequency selection combined with the other approaches.
 \end{itemize}}
 

The paper is structured as follows: in Section~\ref{sec:relatedwork} we discuss the related work and explain the rationale behind the selected mitigation methods; in Section~\ref{sec:WiLabVIsim} we elaborate on the concept of potential interferers, which include both direct and reflected interference; in Section~\ref{Sec:Modeling}, we develop a model to calculate the probability of radar failure, considering a wide number of parameters for all the considered mitigation methods; finally, results are discussed in Section~\ref{Sec:Results} before summarizing our conclusions in Section~\ref{Sec:conclusion}.

\input{RelatedWorkSection}

\section{Distribution of potential interferers}\label{sec:WiLabVIsim}


The approach used in this work is to first define the positions of radars and reciprocal risk of interference, as discussed in this section, and then calculate the actual probability of system failure, as detailed in Section~\ref{Sec:Modeling}. 

We assume realistic positions and movements of the vehicles, by using the open-source road-traffic simulator SUMO \cite{dlr127994}.
More specifically, the scenario runs in SUMO with the given environment (e.g., the highway with three lanes per direction used later in this document) and the given density of vehicles. Vehicles move during the simulation following the road rules and are all equipped with one or more radars with the same settings. 

Starting from the position of vehicles and radars in each snapshot, we then calculate what are called potential interferers and their statistical distribution. The software we developed for the evaluation of the potential interferers is called WiLabVIsim  \cite{BazAtrasc2024} and is openly available.\footnote{The tool is open.source at \href{https://github.com/V2Xgithub/WiLabVIsim}{https://github.com/V2Xgithub/WiLabVIsim}.}

\subsection{Definition of potential interferers} 

\begin{figure}[t]
\centering
\includegraphics[width=0.8\columnwidth]{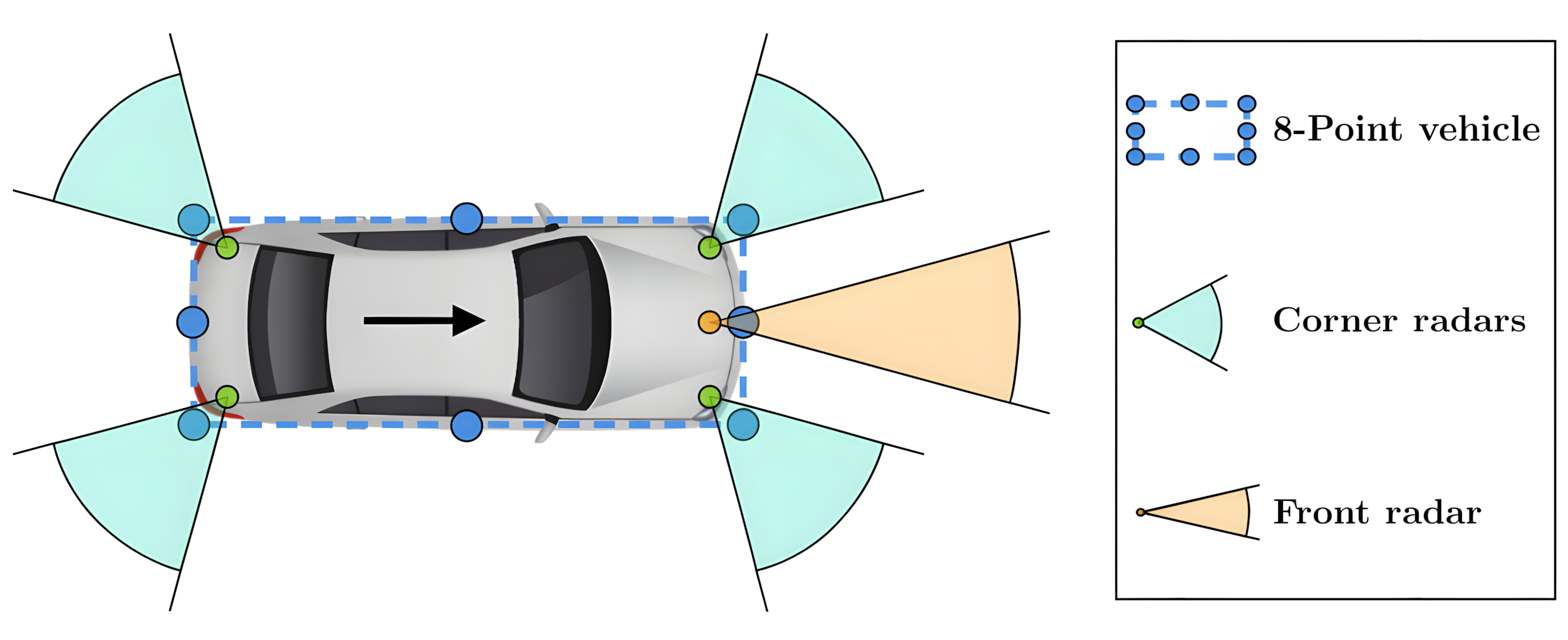}
\caption{Modeling of the vehicle, with either one front radar or four corner radars, and with eight points of reflection for the calculation of reflected interference.}
\label{fig:vehicleRadars}
\vspace{-\baselineskip}
\end{figure}

A radar is called a \textit{potential interferer} of another radar if the signal transmitted by the former arrives inside the \ac{FOV} of the latter one. For the sake of conciseness, hereafter we will use the term \textit{attacker} for the interfering radar and \textit{victim} for the interfered one, although this does not mean that the interference is generated intentionally. 
We consider signals arriving at the receiving radar either directly, i.e., under \ac{LOS} conditions, or after a single reflection, i.e., under \ac{NLOS} conditions. It is in fact assumed that after more than one reflection the signal is very attenuated and becomes negligible 
(please recall that radar systems work around 80~GHz or more, and consider that in our results we assume them at 140 GHz). 
Per each attacker, we consider only the strongest source of interference, which means the direct path if \ac{LOS} conditions between the attacker and victim are met, or the reflected path with the smallest attenuation if \ac{LOS} conditions are not met. If the signal arrives at the victim without reflections, we call the attacker \textit{direct interferer}, otherwise we call it \textit{reflected interferer}. Please remark that the definition of potential interferer does not require to specify the signal characteristics, such as the use of time and frequency resources, but only the position, direction, and \ac{FOV} of the radars, and the presence of obstacles. In our results, the obstacles consist of the other vehicles on the road; each vehicle is represented as a rectangle, and it may obstruct the \ac{LOS} and reflect the received signals. 
With the aim to balance complexity and accuracy, reflections can be generated at one of eight points around the vehicle, following an approach consistent with similar works (see, for example, \cite{6202527}); the eight points are represented in Fig.~\ref{fig:vehicleRadars}. Examples of potential interferers in a specific scenario are provided later in Fig.~\ref{fig:scenario_interferes}.

In order to make direct and reflected interferers comparable, we further introduce the definition of \textit{equivalent distance} $\dref$: given an attacker generating a certain (direct or reflected) interference with a certain interfering power, $\dref$ is the distance at which it would generate the same interfering power if there were no reflections. 
%
This implies that an attacker with \ac{LOS} to the victim and a physical distance $\dref$ is also a potential interferer at the equivalent distance $\dref$.

In the case of reflected interference, denoting as $d_1$ the path length covered by the signal from the attacker to the point of reflection, $d_2$ the one from the point of reflection to the victim, and $\RCS$ the \ac{RCS}, we derive $\dref$ as follows. Let us assume a transmission power $P_\text{t}$, a transmitting antenna gain $G_\text{t}$, a receiving antenna gain $G_\text{r}$, and a wavelength $\lambda$; the power received at distance $d$ can be calculated as
\begin{equation}\label{eq:A1}
    P_\text{r} = \frac{P_\text{t}G_\text{t}G_\text{r}\lambda^2}{(4 \pi)^2 d^2}\;.
\end{equation}
With the same parameters, if the signal is reflected by a surface with an \ac{RCS} $\RCS$ at distance $d_1$ from the transmitter and $d_2$ from the receiver, the power at the receiver can be calculated as 
\begin{equation}\label{eq:A2}
    P_\text{r} = \frac{P_\text{t}G_\text{t}G_\text{r}\lambda^2}{(4 \pi)^2 d_1^2}\;\frac{\sigma}{4 \pi d_2^2}\;.
\end{equation}
Using \eqref{eq:A1} and \eqref{eq:A2}, and replacing $d$ with $\dref$, the equivalent distance at which the victim would receive the same power as it was in \ac{LOS} conditions is 
\begin{equation}\label{eq:disteq}
    \dref = \sqrt{\frac{4 \pi d_1^2 d_2^2}{\RCS}}\;.
\end{equation}

Please observe that the definition of equivalent distance for the reflected interferers depends on the RCS $\sigma$, which in turn depends on the specific surface generating the interference. Although the definition is general (a different RCS could be considered for each reflected interferer), in our results the reflections are always from vehicles and we use the same $\sigma=10$~m$^2$ in all the cases, following the ETSI indications in \cite{ETSI_TR_103_593}.



\subsection{Probability distribution of potential interferers}\label{sec:prodDistribution} 
We denote the \textit{number of potential interferers up to a given equivalent distance $\dmax$} as the number of potential interferers located at an equivalent distance not greater than $\dmax$, thus including signals that may interfere either directly or after one reflection.

By observing all radars across various snapshots of the simulation, it is possible to compute a vector of probabilities representing the likelihood of having a certain number of potential interferers up to a given equivalent distance 
$\dmax$. This vector is hereafter referred to as \textit{probability distribution of potential interferers}  and is denoted as $\PNintVector = \{\PNint{0},\PNint{1},\PNint{2},...\}$. This vector means that $\PNint{0}$ is the probability of having no potential interferers within $\dmax$, $\PNint{1}$ to have one potential interferer within $\dmax$, $\PNint{2}$ to have two potential interferers within $\dmax$, and so on.

Using the probability distribution of potential interferers, we calculate the average probability of system failure while varying the system parameters and applying different interference mitigation methods, as detailed in the following section.

\section{Interference modeling}\label{Sec:Modeling}

In this section, we derive a mathematical model to quantify the impact of interference, based on the statistics of potential interferers described in Section~\ref{sec:WiLabVIsim}. The model assesses how frequently a radar system experiences issues due to interference, assuming the use of \ac{FMCW} with correlated interference and comparing different mitigation methods. In this work, we consider all vehicles to be equipped with either one front-looking (front) radar or four corner-looking (corner) radars as shown in Fig.~\ref{fig:vehicleRadars}, assuming that they operate in separate frequency bands. Interfering radars have the same parametrization and therefore cause correlated interference.

\subsection{Assumptions}

\begin{figure}[t]
\centering
\subfloat[]{%
\centering
         \includegraphics[width=0.9\columnwidth] 
         {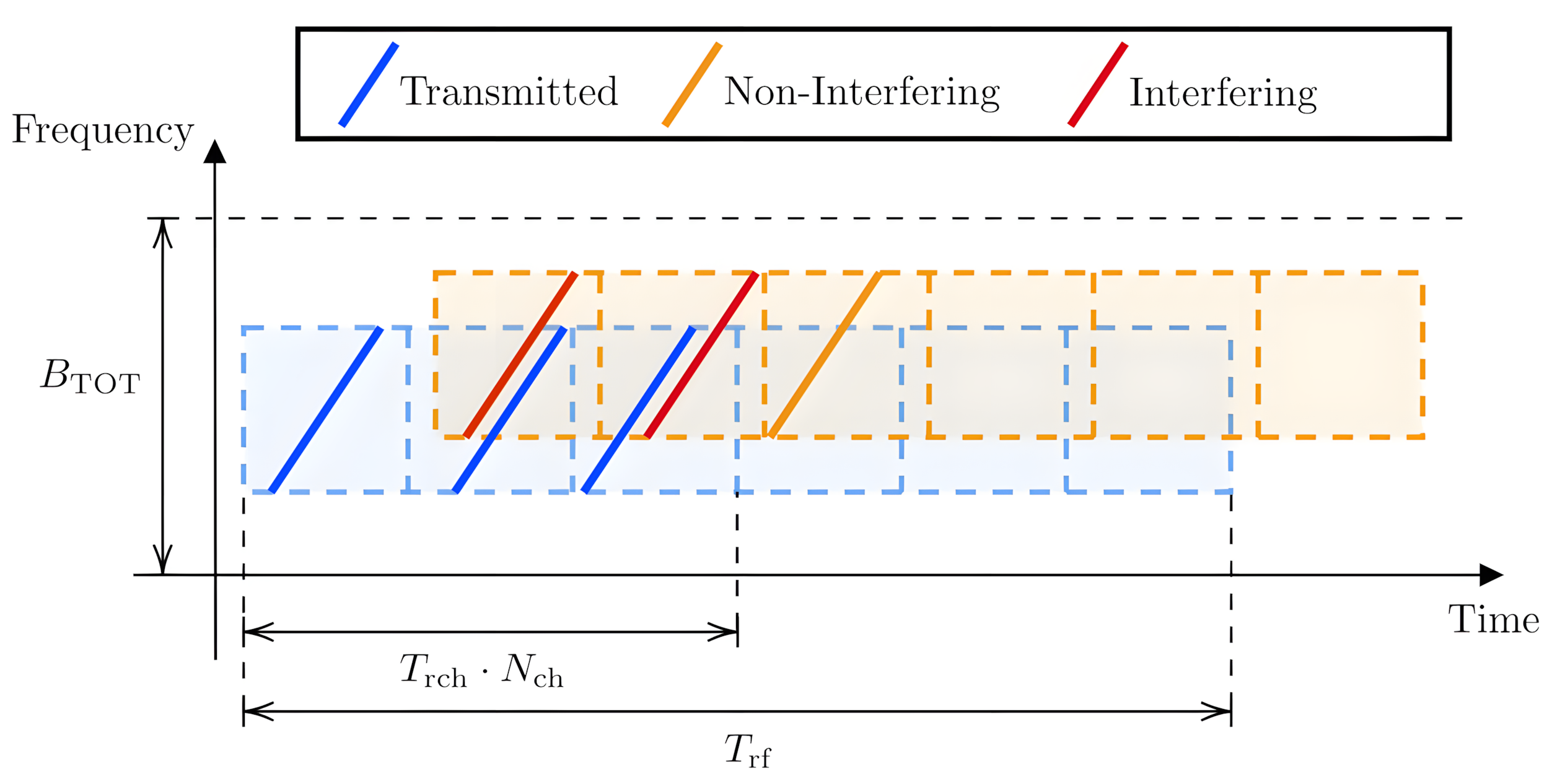}
         \label{fig:frame}    }
~\\
\subfloat[]{%
\centering
         \includegraphics[width=0.9\columnwidth]{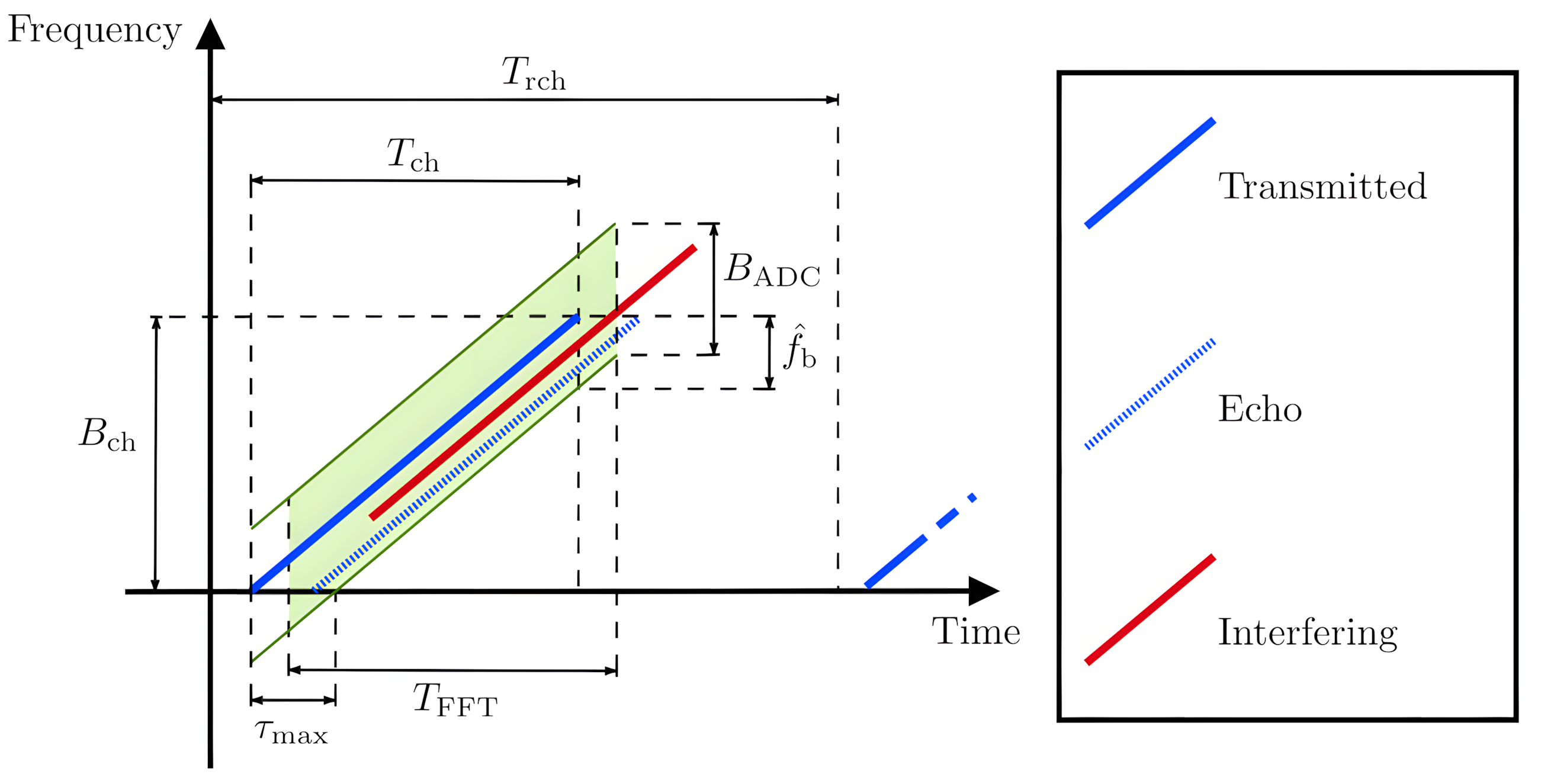}
         \label{fig:chirp}    }
\caption{Representation of the frame (a) and chirp (b).
}\label{fig:frameandchirp}
\vspace{-\baselineskip}
\end{figure}

The \ac{FMCW} frame and chirp structure are illustrated in Fig.~\ref{fig:frameandchirp}. The following assumptions apply: 
\begin{itemize}
    \item The available bandwidth is denoted as $\BTOT$; 
    \item The chirp 
    bandwidth is $\BW$;
    \item The \ac{ADC} 
    bandwidth is $\BADC$;
    \item The maximum beat frequency is $\hat{f}_\text{b}$;
    \item Each chirp is transmitted within a repetition time interval, $\Tchirprep$, which includes the chirp transmission time $\Tchirp$. The starting instant of $\Tchirp$ inside $\Tchirprep$ varies from chirp to chirp to introduce variability. The empty interval also allows the oscillators to be reset; 
    \item The time during which the received signal is down-converted and processed using a double-FFT scheme is denoted as $T_\text{FFT}$, which is assumed to be numerically equal to $\Tchirp$;
    \item The maximum one-way delay for detecting a reflected chirp is given by $\tau_\text{max}=\Tchirp \hat{f}_\text{b}/\BW$, determining the maximum detectable target distance as $\tau_\text{max} \cdot c$, where $c$ is the speed of light;
    \item Each radar transmits a sequence of chirps, referred to as a \textit{frame}, with an empty interval at the end. The number of chirps per frame is denoted as $\nch$. The \textit{duty cycle}, $\duty$, represents the ratio of active time to the total frame duration. The total frame duration, including active and inactive states, is denoted as $\Tframe$ and can be calculated as 
    \begin{equation}\label{eq:Tframe}
        \Tframe=\Tchirprep \cdot \nch / \duty \;.
    \end{equation}
    We assume that $\duty$ is set such that $\Tframe$ is a multiple of $\Tchirprep$;
    \item A chirp transmitted by the victim is assumed to be a \underline{chirp colliding} with that of an attacker if all the following conditions are met: (i) the attacker is a potential interferer within the maximum defined equivalent distance; (ii) there is an overlap in the frequency domain between the victim's and attacker's signal by at least $\xf \in [0,1]$ (i.e., if the overlap is less than $\xf$, the impact of the interfering signal is considered negligible); and (iii) the chirp of the attacker, once down-converted, enters the \ac{ADC} bandwidth of the victim;
    \item A \underline{frame loss} for the victim is assumed if the number of chirps collided with any of the attackers is equal to or exceeds a minimum value denoted as $\kch$;
    \item The radar system experiences a \underline{system failure} if $\Mf$ consecutive frames are lost.
\end{itemize}

\subsection{Interference mitigation modeling}

We consider and compare the following cases:
\begin{enumerate}
    \item A \textit{baseline} scenario in which each radar is randomly assigned an initial starting frequency that remains fixed;
    \item A \textit{frame-by-frame} frequency hopping approach, where each radar selects randomly a new starting frequency for every frame;
    \item A \textit{chirp-by-chirp} frequency hopping, where each radar randomly selects a new starting frequency for every chirp;
    \item The integration of the \textit{compass} method with either frame-by-frame or chirp-by-chirp frequency hopping, using different configurations.
\end{enumerate}

\subsubsection{Baseline}\label{subsec:benchmark}

As a baseline, we assume that the starting frequency of each radar is randomly selected within $\BTOT-\BW$  according to a uniform distribution. Since there is no correlation between the use of frequency and time resources, calculations are first performed separately in the frequency and time domains.

\textit{In the frequency domain,} the probability that an attacker's signal overlaps with that of the victim by at least a fraction $\xf$ can be calculated as
\begin{align}\label{eq:Pfo}
    \Pfo &= \frac{2\cdot(1-\xf)\cdot\BW}{\BTOT-\BW} \cdot \frac{\BTOT-\BW-(1-\xf)\cdot\BW/2}{\BTOT-\BW}\;.
\end{align}
The derivation of \eqref{eq:Pfo} is detailed in Appendix~A. 


Recalling that, in the baseline approach, the starting frequency is not modified during operation, the probability distribution of potential interferers and the probability calculated in \eqref{eq:Pfo} can be used to evaluate the probability of actually having no attackers overlapping in the frequency domain, one attacker overlapping, and so on. Such vector of probabilities, denoted as $\PNintRevVector = \{\PNintRev{0},\PNintRev{1},\PNintRev{2},...\}$, has elements that can be calculated starting from the vector $\PNintVector = \{\PNint{0},\PNint{1},\PNint{2},...\}$ as follows:
\begin{align}\label{eq:PNintRev}
    \PNintRev{n} = \sum_{j=n}^{\infty} \PNint{j} \binom{j}{n} \left(\Pfo\right)^n \left(1-\Pfo\right)^{(j-n)}\;.
\end{align}
The derivation of \eqref{eq:PNintRev} is detailed in Appendix~B. 

\textit{In the time domain}, let us start considering a single attacker. Depending on the instants when the frames transmitted by the attacker and the victim start, there can be a varying number of overlapping chirp repetition intervals. Every time these chirp repetition intervals overlap, there is a certain probability that the chirp sent by the attacker actually causes a collision. Assuming that such an overlap occurs between the chirp repetition intervals used by the attacker and the victim, the probability that the signal from the attacker causes a collision is
\begin{equation}\label{eq:Ptchirp}
    \Ptchirp=\frac{\Tchirp}{\Tchirprep} \cdot \frac{\BADC}{\BW}
\end{equation}
where the left term accounts for the variability in the starting point of the chirp within the chirp repetition interval, while the right term considers that the attacker's signal is filtered out if it does not fall within the ADC bandwidth. Then, considering the possible starting instants of the frames, the probability that at least $\kch$ out of the $\nch$ chirps in the attacker's frame collide can be approximated as
\begin{equation}\label{eq:Ptb}
\PtB = \sum_{z=\kch}^{\nch}\frac{2}{\nch/\duty}\left(1-\Psi(z) \right)
\end{equation}
where
\begin{equation}\label{eq:Ptb2}
\Psi(z)=\sum_{j=0}^{\kch-1} \binom{z}{j} \Ptchirp^j (1-\Ptchirp)^{z-j}\;.
\end{equation}
The summation in \eqref{eq:Ptb} considers the possible number of chirp repetition intervals that overlap and the summation in  \eqref{eq:Ptb2} the number of chirps within the overlapping chirp repetition intervals that actually collide. The derivation of \eqref{eq:Ptb} is detailed in Appendix~C.

 \rev{If there are N attackers overlapping in the frequency domain, and assuming that a collision occurs if there is a collision with any of the attackers, the probability of frame loss becomes} 
\rev{\begin{equation}\label{eq:PfailB}
\PcollB{N} = 1-(1-\PtB)^N\;.
\end{equation}}
\rev{Finally, let us consider $M$ consecutive transmissions and assume that the overlapping radars over such transmissions remain the same. For a given $M$, we assume that consecutive frames are lost independently, so that the probability to have $M$ consecutive losses when there are $N$ interferers is  $\left(\PcollB{N}\right)^M$. Hence, accounting for the statistic of the number of actual interferers, the probability of system failure is obtained as}
\rev{\begin{equation}\label{eq:pfailC}
\PfailB = \sum_{n=1}^{\infty} \left[\PNintRev{n}\left(\PcollB{n}\right)^M\right] \;.
\end{equation}}

\subsubsection{Frame-by-frame frequency hopping}\label{subsec:frame_by_frame}

When a frame-by-frame frequency hopping strategy is used, the starting frequency of the chirp is randomly changed at the beginning of every frame. The selection is performed randomly within $\BTOT-\BW$ with uniform distribution. 

Recalling that: (i) the probability that the selection of a random starting frequency causes an overlap of at least $\xf$ with the probability calculated in \eqref{eq:Pfo}; and (ii) the probability that, given one attacker, there are $\kch$ or more chirps colliding in the time domain can be calculated as in \eqref{eq:Ptb}, in the presence of a single attacker the probability that a frame is lost can be calculated as 
\begin{equation}\label{eq:pfailFone}
\PcollF{1}=\Pfo \PtB\;.
\end{equation}
Thus, in the presence of $N$ attackers that independently select the starting frequency and assuming that there is no collision if there is no collision with any of the attackers, the probability that a frame is lost can be calculated  as
\begin{equation}\label{eq:PcollF}
\PcollF{N}=1-\left(1-\Pfo \PtB\right)^N\;.
\end{equation}

Introducing the distribution of potential interferers, assuming \rev{that the number of potential interferers do not change during $M$ consecutive transmissions, and assuming that there is} no correlation between consecutive frame losses \rev{when looking at frequency shifts of the frames and time shifts of the chirps}, the probability of system failure becomes 
\rev{\begin{equation}\label{eq:pfailF}
\PfailF = \sum_{n=1}^{\infty}\left[\PNint{n} \left(\PcollF{n}\right)^M \right] \;.
\end{equation}}

Please note that the frame-by-frame frequency hopping strategy becomes equivalent to the baseline if the total bandwidth is equal to the chirp bandwidth or, more precisely, if $\Pfo=1$.

\subsubsection{Chirp-by-chirp frequency hopping}\label{subsec:chirp_by_chirp}

When a chirp-by-chirp frequency hopping strategy is used, the starting frequency of the chirp is randomly changed at the beginning of every chirp. The selection is performed randomly within $\BTOT-\BW$ with uniform distribution. In this case, assuming a single attacker, the probability of losing the frame is similar to the one calculated in \eqref{eq:Ptb}, where the probability that the generic chirp is lost is not only dependent on the variability in the time domain, but also in the frequency domain. As a consequence, the probability of losing the frame in the presence of a single attacker becomes
\begin{equation}\label{eqPcollC}
\PcollC{1} = \sum_{z=\kch}^{\nch}\frac{2}{\nch/\duty}\left(1-\chi(z)\right)
\end{equation}
where
\begin{equation}\label{eqPcollC2}
\chi(z) = \sum_{j=0}^{\kch-1} \binom{z}{j} (\Pfo\Ptchirp)^j (1-\Pfo\Ptchirp)^{z-j} \;. 
\end{equation}
In the presence of $N$ attackers that independently select the starting frequency of each chirp and assuming 
that there is no collision if there is no collision with any of the attackers, the probability that a frame is lost can be calculated as
\begin{equation}\label{eq:PcollC}
\PcollC{N}=1-\left(1-\PcollC{1}\right)^N\;.
\end{equation}
Introducing the distribution of potential interferers, assuming \rev{that the number of potential interferers remains constant over $M$ consecutive transmissions and assuming that there is} no correlation between consecutive frame losses \rev{when looking at frequency and time shifts of the chirps},  
the probability of system failure becomes 
\rev{\begin{equation}\label{eq:pfailC}
\PfailC = \sum_{n=1}^{\infty} \left[\PNint{n}\left(\PcollC{n}\right)^M\right] \;.
\end{equation}}

Please note that also the chirp-by-chirp frequency hopping strategy becomes equivalent to the baseline if the total bandwidth is equal to the chirp bandwidth or, more precisely, if $\Pfo=1$.

\subsubsection{Use of compass}\label{subsec:compass}

The compass mitigation strategy assumes that the direction of the radar is used to determine the starting frequency of the chirp. If the total spectrum is divided in channels and the direction of the radar is used to select the channel but not the starting frequency within the channel, then this approach can be combined with frame-by-frame and chirp-by-chirp. Specifically, the round angle is divided into a number of sectors equal to the number of channels, and each sector is associated with a channel; then, the sector in which the radar direction falls determines the channel. On the one hand, the use of compass statistically reduces the number of potential interferers, but on the other hand it reduces the bandwidth available for the randomization of the starting frequency of the chirps. 

In the case of use of compass with frame-by-frame or chirp-by-chirp, the same equations \eqref{eq:pfailF} and \eqref{eq:pfailC} can be used, respectively, where: (i) the distribution of potential interferers is re-calculated taking into account the direction of the attacker; and (ii) the total bandwidth is reduced by a factor equal to the number of sectors. 

\begin{figure*}[t]
\centering
\subfloat[]{%
\centering
         \includegraphics[width=0.55\columnwidth]{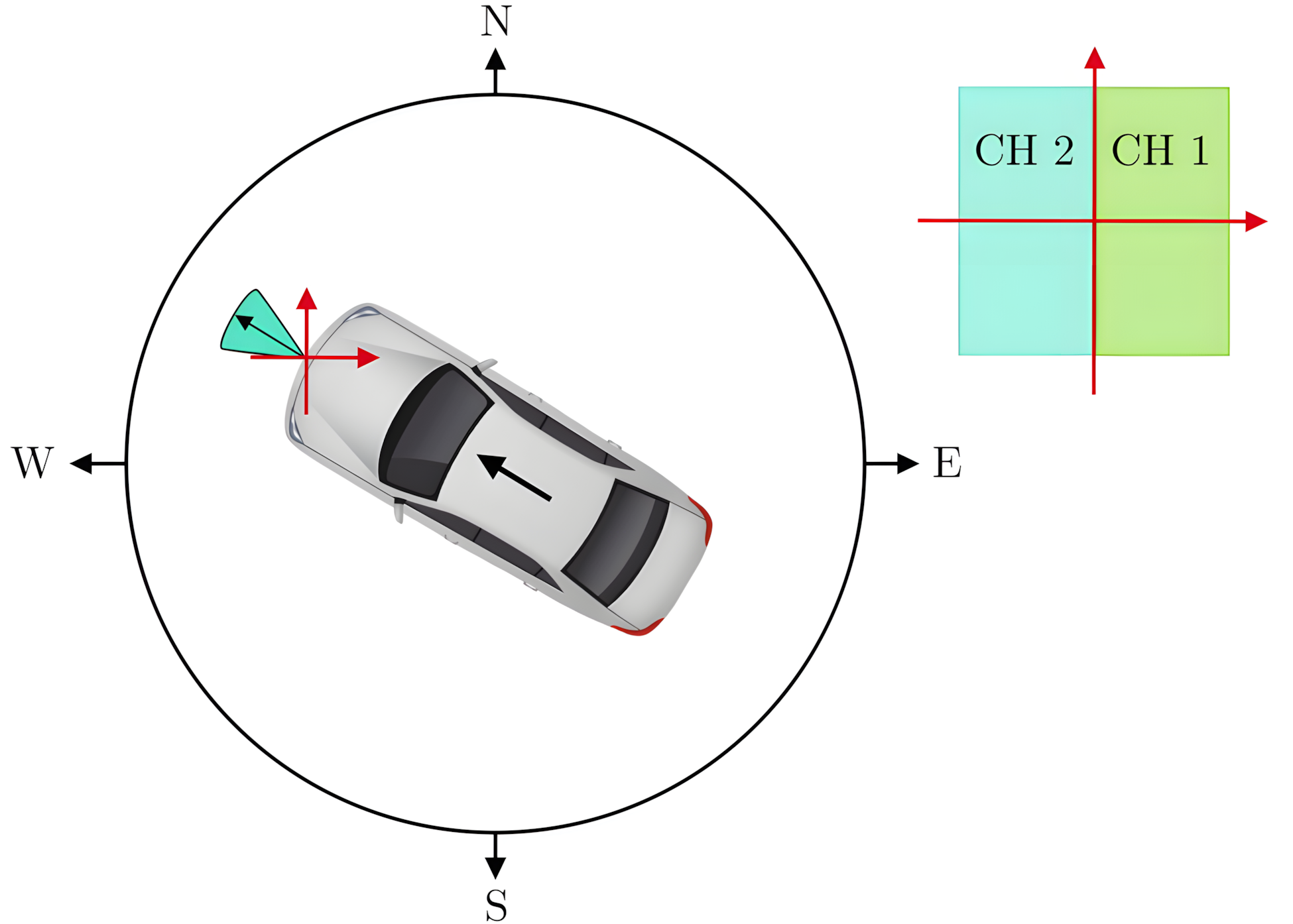}
         \label{fig:compass_a}    }
~
\subfloat[]{%
\centering
         \includegraphics[width=0.55\columnwidth]{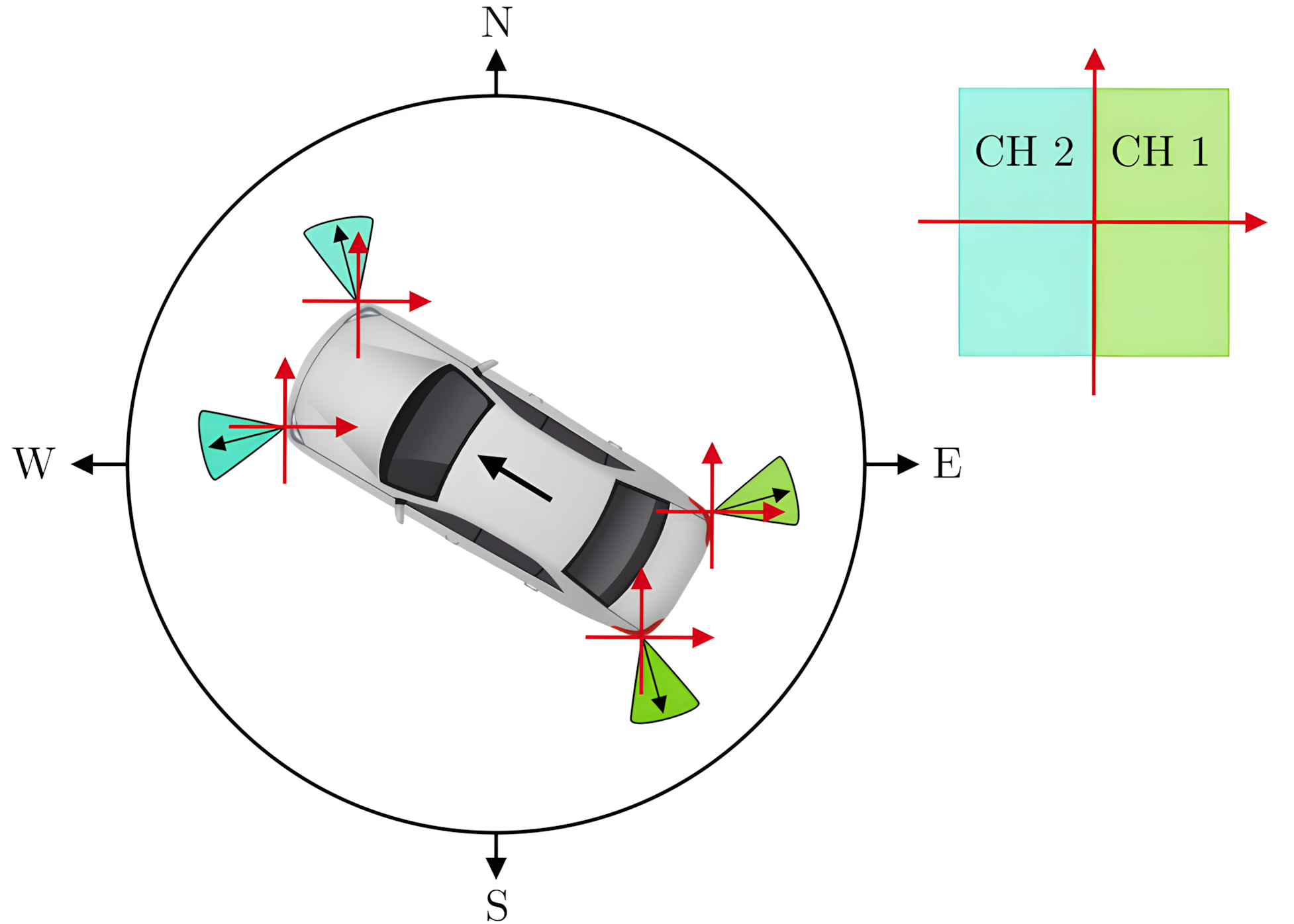}
         \label{fig:compass_b}    }
~\subfloat[]{%
\centering
         \includegraphics[width=0.55\columnwidth]{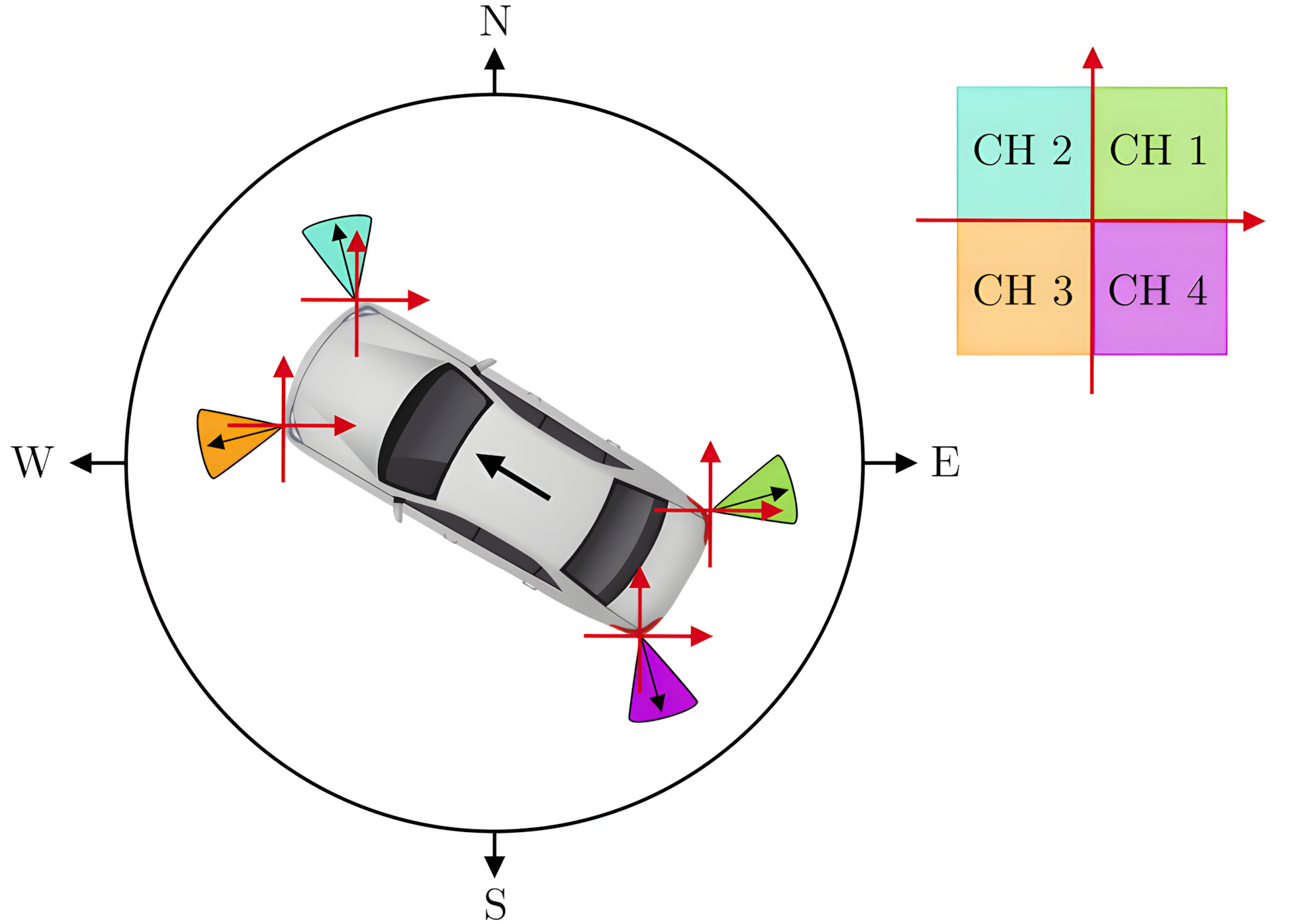}
         \label{fig:compass_c}    }
\\[1mm]
\caption{Compass configurations. Front radar with two sectors (a), corner radar with two sectors (b), and corner radar with four sectors (c).
}\label{fig:compass}
\end{figure*}

In the numerical results, the different configurations exemplified in Fig.~\ref{fig:compass} are assumed. Specifically, in the case of front radar, two channels are assumed (Fig.~\ref{fig:compass_a}), while in the case of corner radars, either two channels (Fig.~\ref{fig:compass_b}) or four channels (Fig.~\ref{fig:compass_c}) are assumed. In the case of corner radars with two sectors, the channelization is performed to make the corner radars in the front of the vehicle on a separate channel compared to those in the back, as this was found slightly more effective than separating those on the one side to those on the other side.

\input{TableSettings_1}
\input{TableSettings_2}


\section{Results}\label{Sec:Results}

The results presented in this section are obtained using the detailed models and, \rev{where not differently specified,} the radar parameters outlined in Tables~\ref{tab:settings_1} and~\ref{tab:settings_2}. These parameters are those being used in ETSI TR 104 054, which are intended for the 140~GHz band. 
\rev{The derived conclusions are anyway not limited to these specific settings, as will be demonstrated in Section~\ref{sec:additional_settings}.} After showing statistics on the number of potential interferers, the performance of the proposed techniques is evaluated in terms of the average time between two system failures, calculated as:
\begin{equation}
    \Tfail = \Tframe \cdot \frac{1}{\PfailX}
\end{equation}
where $\PfailX$ is the failure probability using either of the mitigation methods, i.e., adopting \eqref{eq:PfailB} for the baseline, \eqref{eq:pfailF} for the frame-by-frame, and \eqref{eq:pfailC} for the chirp-by-chirp. When the compass method is used, the same equations apply, but starting from different distributions of the number of potential interferers and different bandwidths, as explained in Section~\ref{subsec:compass}.

To provide a meaningful reference, the average time between system failures is compared to the average use of a car over a week and a year, assuming 8 hours and 22 minutes spent in the car per week.\footnote{Statistics are taken from \href{https://wpde.com/news/auto-matters/americans-spend-18-days-in-their-car-per-year-forge-close-bonds-with-a-vehicle-study}{https://wpde.com/news/auto-matters/americans- spend-18-days-in-their-car-per-year-forge-close-bonds-with-a-vehicle-study}\;.} Furthermore, by examining the trend of $T_\text{fail}$ as the total bandwidth $B_\text{TOT}$ varies, it is possible to assess the effectiveness of interference mitigation techniques combined with the benefit of using a larger bandwidth at 140~GHz.

\begin{figure*}
    \centering
    \includegraphics[width=1\textwidth]{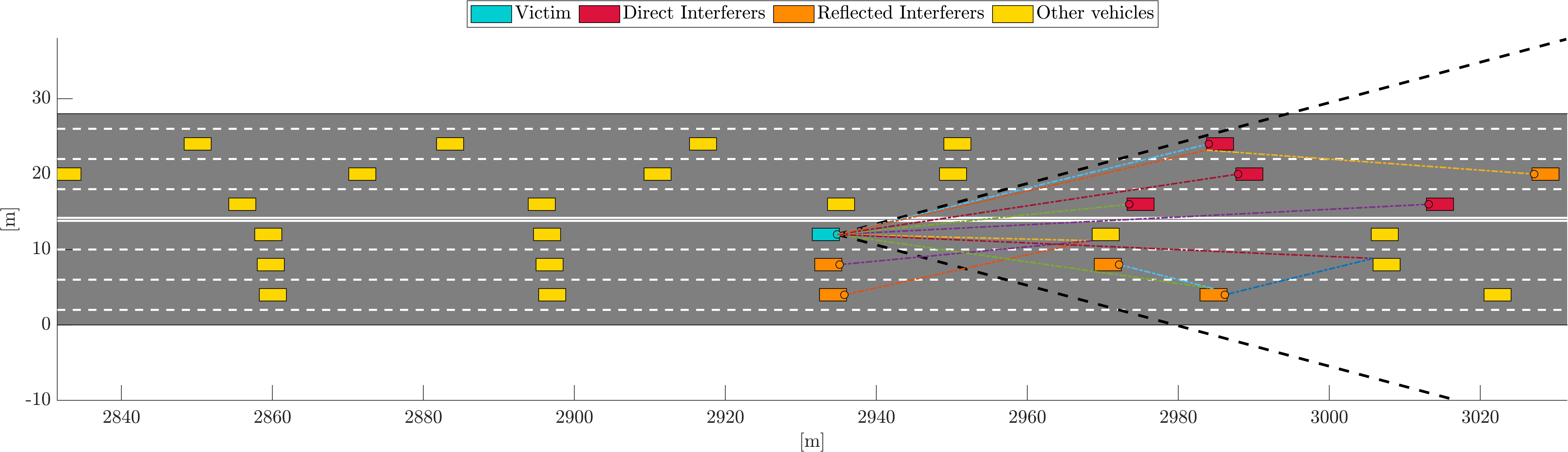}
    \caption{Example of a radar affected by direct and reflected potential interferers in a scenario with 150 vehicles per km, where each vehicle is equipped with a front radar.}
    \label{fig:scenario_interferes}
\end{figure*}

\subsection{Simulation settings}
The scenario under consideration is a highway with 3 lanes per direction and a total length of 8 km. The evaluation covers three levels of congestion: 60, 150, and 270~veh/km with uniform traffic in the two directions,  
where each vehicle is equipped with one front radar or four corner radars. 
Each radar is assumed to have an antenna with constant gain within the given FOV and zero outside; when the corner radars are assumed, the aperture guarantees no overlapping between the FOV of the four radars. 
It can be observed that 60, 150, and 270 veh/km correspond to one vehicle every 100, 40, and 22~m in each lane on average, respectively. As an example, a snapshot (i.e., a single instant) of the scenario simulated using SUMO is shown in Fig.~\ref{fig:scenario_interferes} in the case with 150~veh/km; in the figure, all vehicles are equipped with the front radar, the vehicle with the victim radar is colored in light blue, those with the direct potential interferers in red, those with the reflected potential interferers in orange, and those with radars that are not potential interferers of the victim in yellow; black dashed lines indicate the \ac{FOV} of the victim radar and dashed-dotted lines with different colors the path that is covered by the signals sent from the potentially interfering radars to the victim radar.

A radar is considered an attacker if and only if the received power of the potentially interfering signal is above a minimum threshold, corresponding to a minimum \ac{INR}.  The maximum equivalent distance can be computed as 
\begin{equation}
\dmax = \dfrac{\light}{4 \pi \fzero} \cdot 10^{{(\EIRPdBm-30 + \Gr - 10 \log_{10}(\Boltzmann*\Tzero*\BADC)-\NF-\INRmindB)}/{20}}
\end{equation}
where $\light$ is the speed of light, $\fzero$ is the carrier frequency, $\EIRPdBm$ is the \ac{EIRP} expressed in dBm, $\Gr$ is the gain of the receiving antenna expressed in dB, $\Boltzmann$ is the Boltzmann constant, $\Tzero$ is the reference temperature of 290~K, $\NF$ is the receiver noise figure expressed in dB, and $\INRmindB$ is the minimum \ac{INR} expressed in dB. Table~\ref{tab:settings_2} summarizes all the parameters used to determine the maximum distance at which an interfering signal can impact the victim's radar.
With the given parameters, we computed a maximum equivalent distance of 2.7~km and 120~m for front and corner radars, respectively.

As shown in Table~\ref{tab:settings_1}, the analysis assumes, when not differently specified, a pessimistic radar system where a frame is deemed lost if at least 5\% of the chirps are lost and a system failure occurs after 3~consecutive frame losses. The approach was validated using Monte Carlo simulations of \ac{FMCW} receivers employing \ac{CA-CFAR} algorithms. 



\subsection{Potential interferers}\label{sec:symmPotInterf}

\begin{figure*}[t]
\centering
     \subfloat[]{
         \includegraphics[width=0.45\textwidth,draft=false]{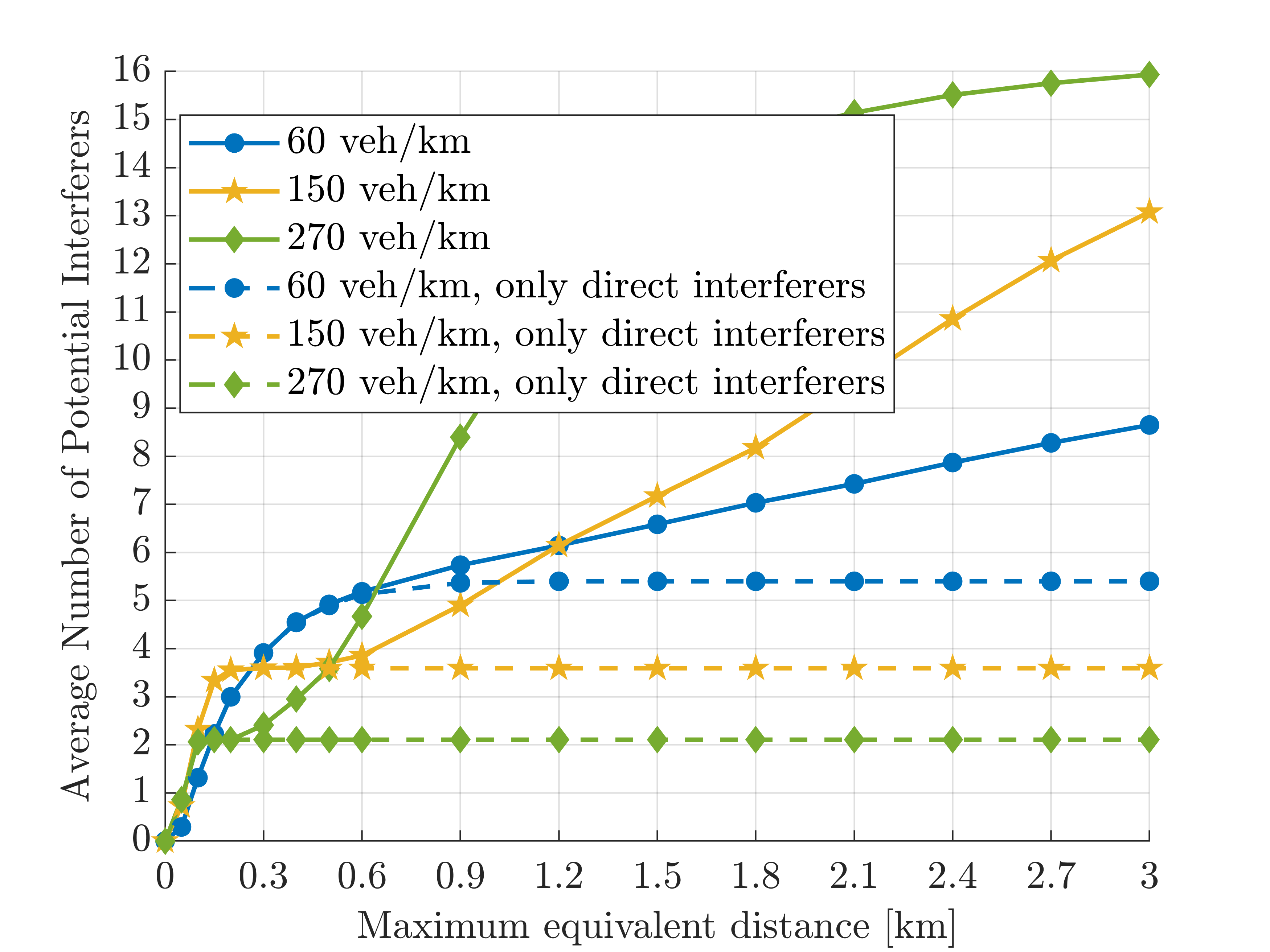}
          \label{fig:front_direct_equivalent_symmetric}
      }\subfloat[]{
         \includegraphics[width=0.45\textwidth,draft=false]{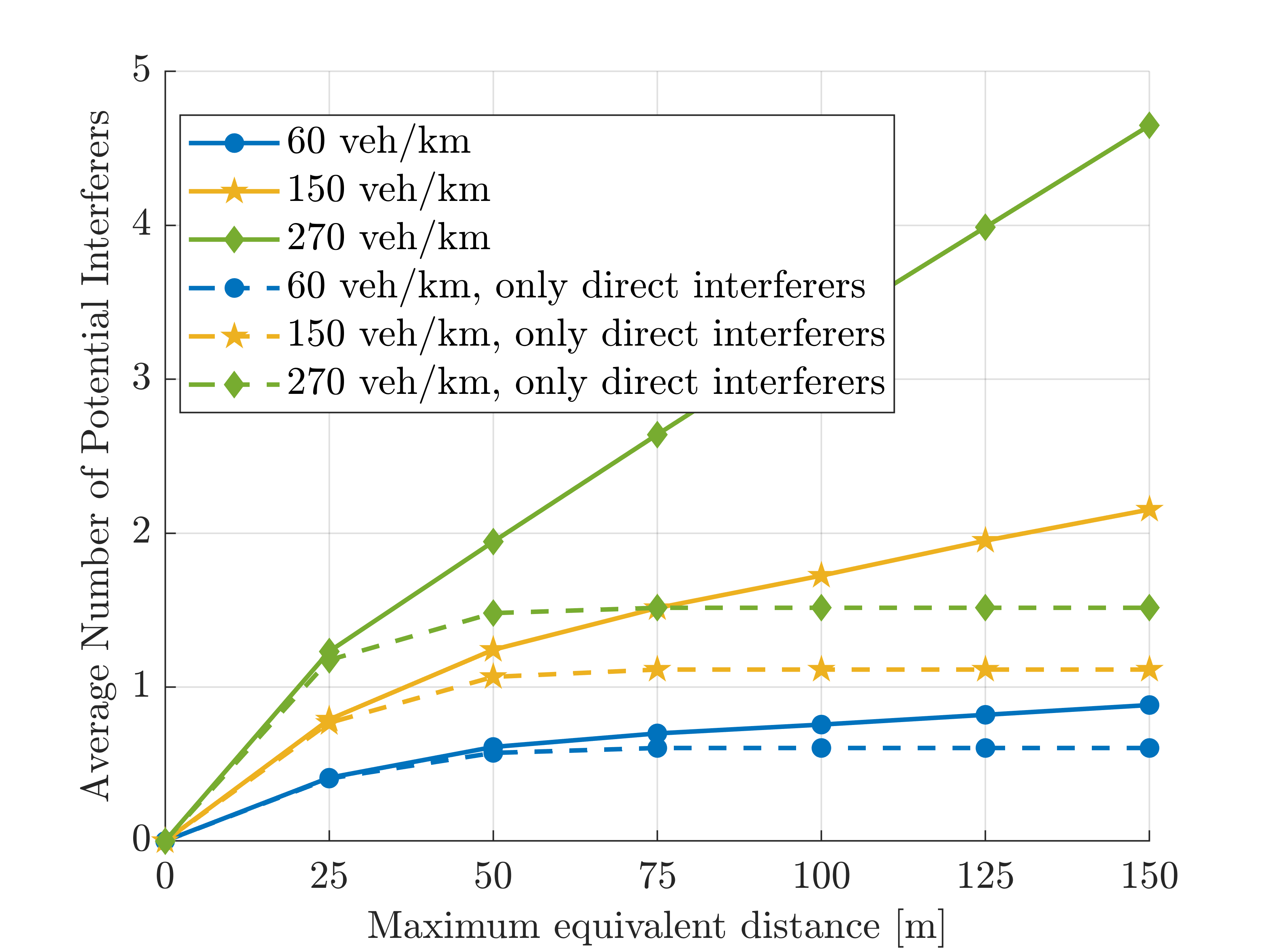}
          \label{fig:corner_direct_equivalent_symmetric}
     }\\[1mm]
        \caption{Average number of potential interferers varying the maximum equivalent distance, including or excluding the reflected interference. Front (a) or corner (b) radars.}
        \label{fig:interferers_equivalent_symmetric}
\end{figure*}

The previous assumptions allow us to investigate the number of potential interferers without restricting the study to the maximum equivalent path length derived in Table~\ref{tab:settings_2}. Fig.~\ref{fig:interferers_equivalent_symmetric} illustrates the average number of potential interferers as the maximum equivalent distance varies across different traffic densities, representing low, medium, and high congestion levels on the highway. Per each density and both types of radars, the number of potential interferers increases rapidly with small values of the maximum equivalent distance and then the increase becomes less steep. Furthermore, comparing the solid and dashed curves it appears clear that for small values of the maximum equivalent distance the number of potential interferers increases mainly due to direct interferes, which then have a very limited additional contribution for large values of the maximum equivalent distance; this trend is due to the following considerations: (i) the power attenuation due to the reflection makes negligible the probability of having a potential reflected interferer at a short equivalent distance; (ii) when the source of the direct interferer is far from the victim radar, there is a very high probability that there is another vehicle in the between acting as obstacle and therefore blocking the \ac{LOS}; (iii) in the case of corner radars, the maximum distance of a direct interferer is limited by the geometry of the scenario, since in the worst case the attacker and victim are located in the rightmost lanes of the two directions.

\begin{figure*}[t]
     \centering
     \subfloat[]{
         \includegraphics[width=0.45\textwidth,draft=false]{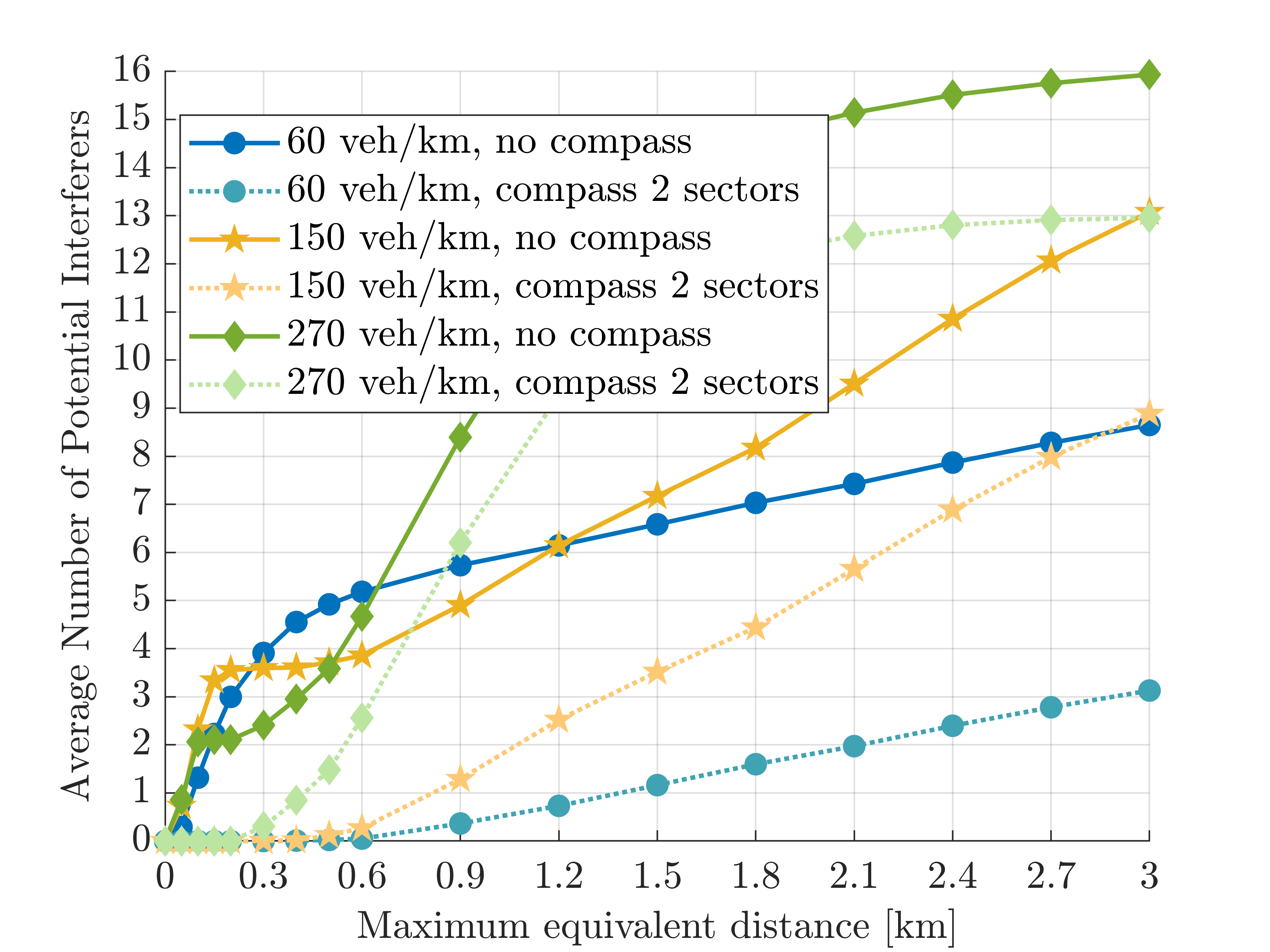}
          \label{fig:front_direct_equivalent_compass_symmetric}
     }\subfloat[]{
         \includegraphics[width=0.45\textwidth,draft=false]{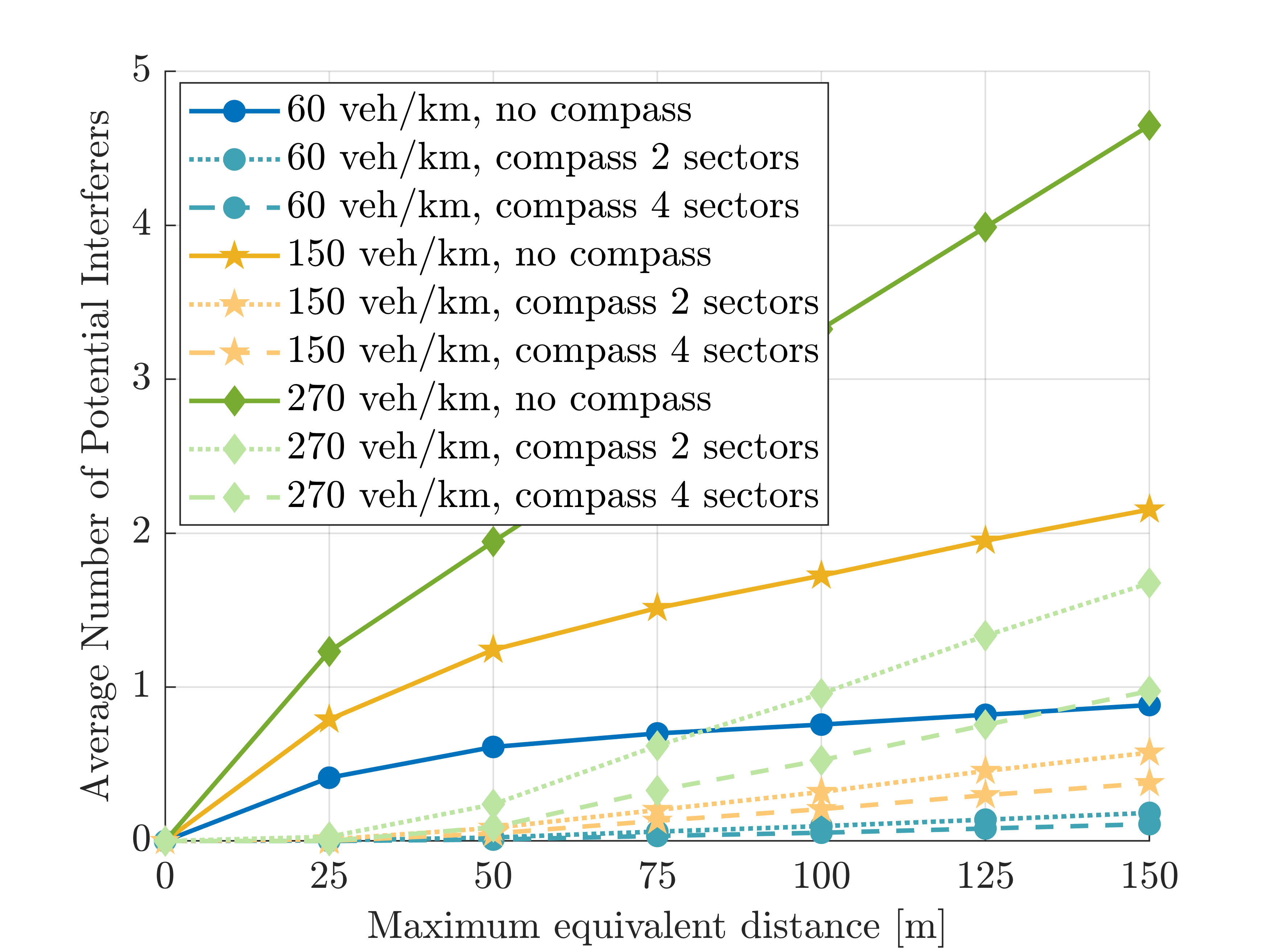}
          \label{fig:corner_direct_equivalent_compass_symmetric}
     }\\[1mm]
        \caption{Average number of potential interferes, with and without compass, varying the the maximum equivalent distance. Front (a) or corner (b) radars.}
        \label{fig:interferers_equivalent_compass_symmetric}
\end{figure*}

Another relevant consideration derives by comparing the curves of Fig.~\ref{fig:interferers_equivalent_symmetric} corresponding to different vehicle densities. In the case of front radars, a larger density increases the overall number of potential interferers but reduces the number of direct potential interferers. As the density of vehicles increases, in fact, direct interference decreases due to more vehicles acting as obstacles. 
Differently, for corner radars, similar trends are observed when analyzing the total interference and only direct interference; in this case, the effect of the vehicles acting as obstacles is limited due to the few lanes between the attackers and the victims and a larger density causes an increase of the number of potential direct interferers.

The statistics of potential interferers are further analyzed in Fig.~\ref{fig:interferers_equivalent_compass_symmetric}, which again shows the number of potential interferers varying the maximum equivalent distance, but includes the case where the compass method is added. What can be observed is that compass is particularly effective in scenarios with many direct interferers, which means that its effectiveness decreases as the traffic density increases. Similarly, its effectiveness is significant for small equivalent distances and therefore appears efficient to reduce the impact of strong interferers. In the case of corner radars, it is worth observing that implementing four instead of two sectors does not lead to a significant reduction of the potential interferers.


\begin{figure*}[t]
     \centering
     \subfloat[]{
         \includegraphics[width=0.45\textwidth,draft=false]{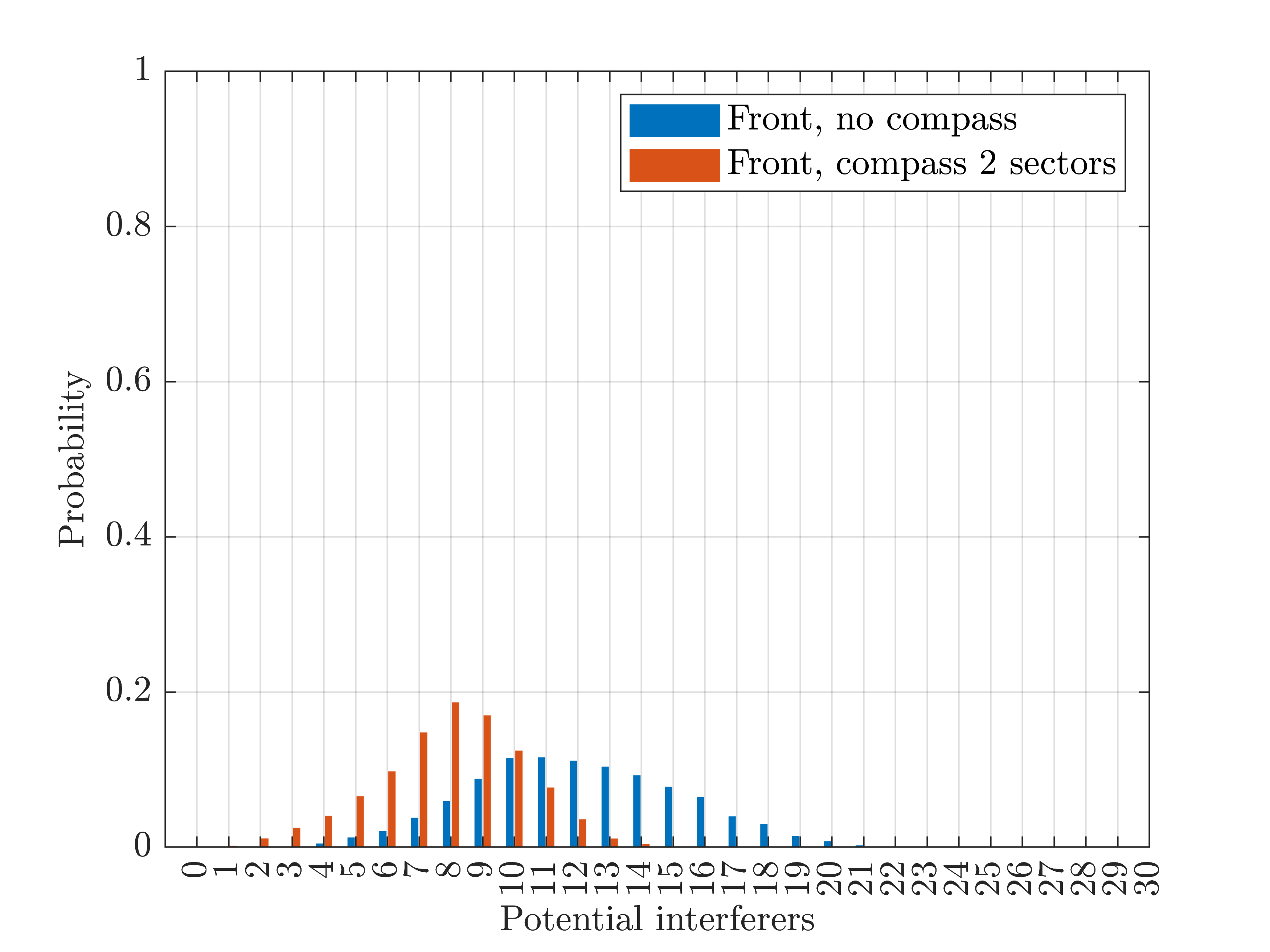}
          \label{fig:front_bar_symmetric}
     }\subfloat[]{
         \includegraphics[width=0.45\textwidth,draft=false]{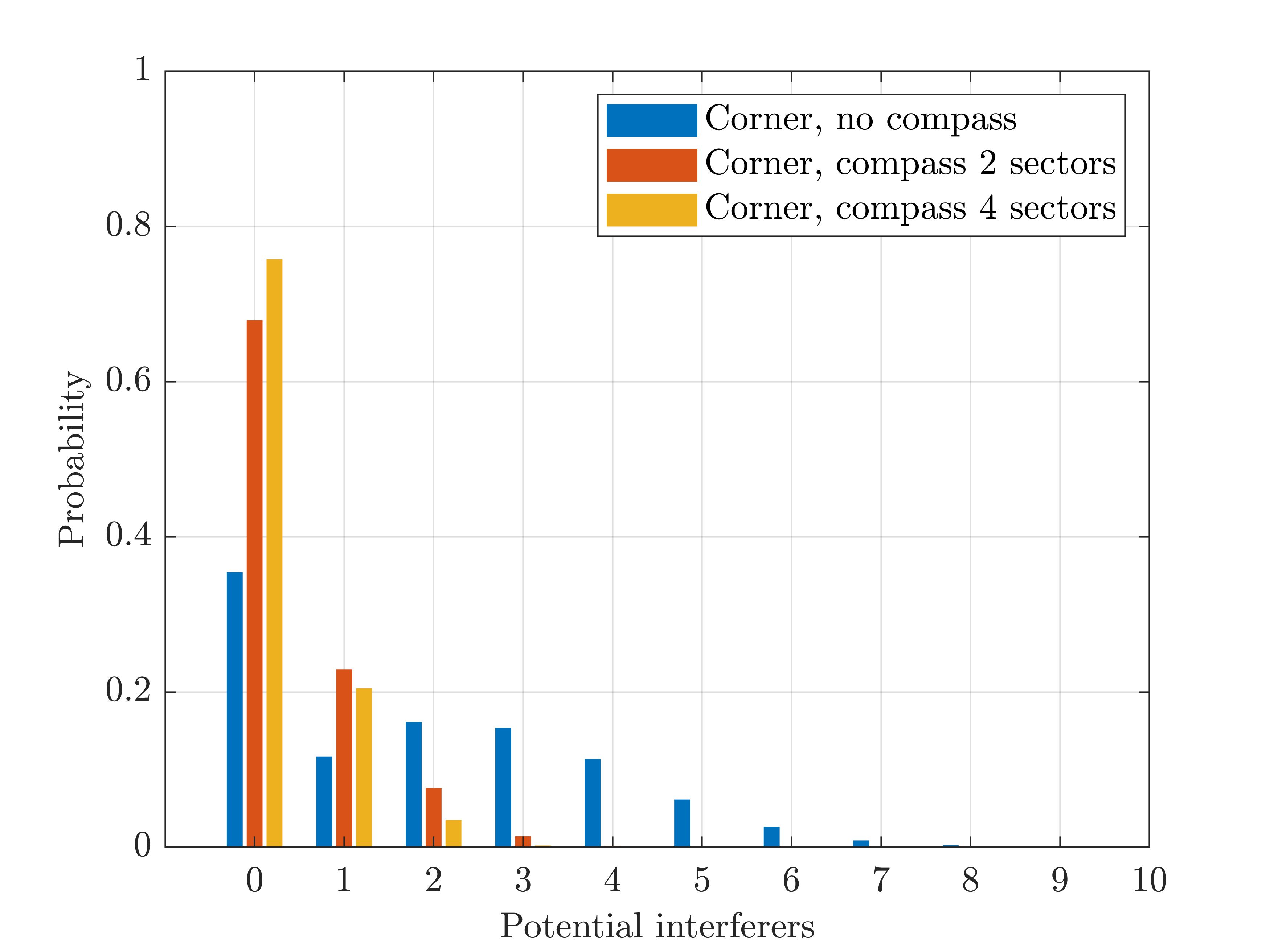}
          \label{fig:corner_bar_symmetric}
     }\\[1mm]
        \caption{Distribution of interferes with 150 veh/km. Front (a) or corner (b) radars.}
        \label{fig:interferers_bar_symmetric}
\end{figure*}

\begin{figure*}[t]
      \centering
     \subfloat[]{
         \includegraphics[width=0.45\textwidth,draft=false]
         {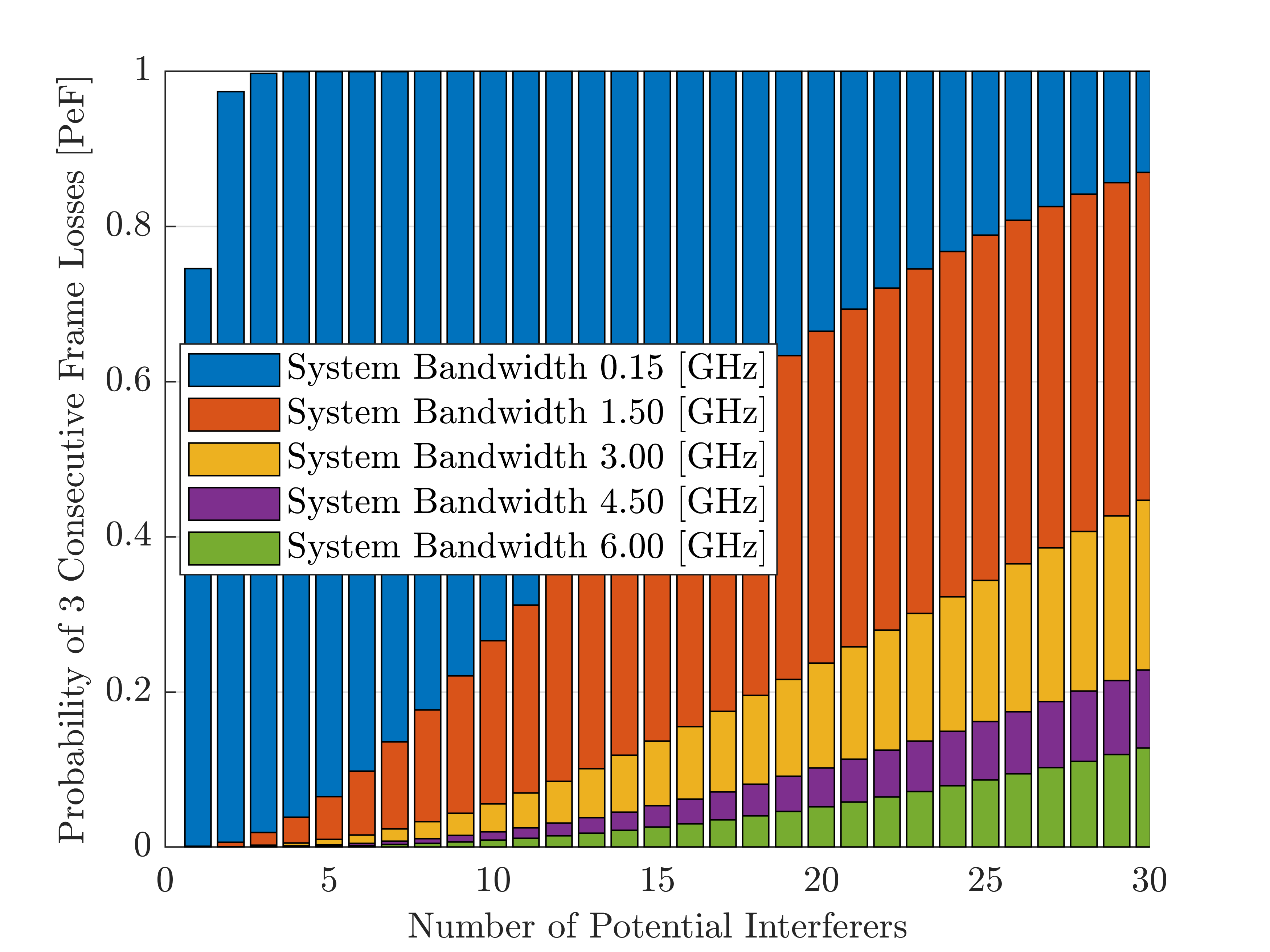}
          \label{fig:front_PlotPeF}
     }
          \subfloat[]{
         \includegraphics[width=0.45\textwidth,draft=false]
         {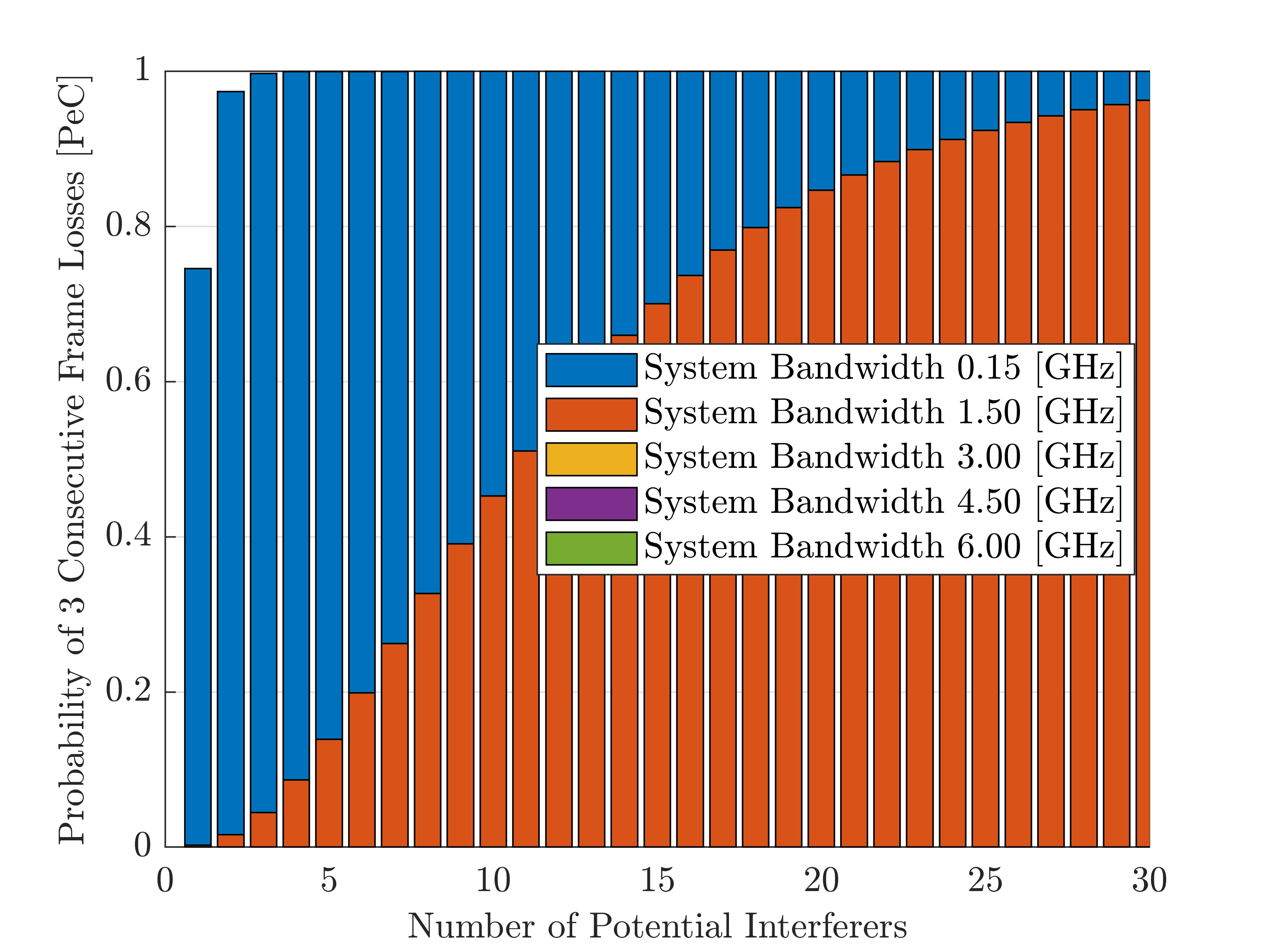}
          \label{fig:front_PlotPeC}
     }
        \caption{\rev{Probability to have $M=3$ consecutive frame losses given that there are $N$ potential interferers, varying $N$, for various values of the total available bandwidth. (a) Applying frame-by-frame. (b) Applying chirp-by-chirp.}}
        \label{fig:front_PlotPe}
\end{figure*}

Restricting the study to the maximum equivalent distance indicated in Table~\ref{tab:settings_2} and to the scenario with 150 veh/km, Fig.~\ref{fig:interferers_bar_symmetric} illustrates the probability distribution of the number of potential interferers with and without compass. As observable, in the case of front radars, where the maximum equivalent distance is of 2.7~km, there can be more than 20 potential interferers in the worst case, with a negligible probability of having one or no potential interferers. The compass method effectively reduces the probability of having large number of potential interferers, which however remains often in the order of 7 to 9. When looking at the corner radars, the situation is less critical, both because of the geometry of the scenario and because of the lower maximum equivalent distance due to the lower transmission power; in this case, without compass the maximum number of potential interferers is lower than 10, and they are reduced to a few adding compass.

\subsection{Impact of the mitigation methods}\label{sec:symmMitigation}

\begin{figure*}[t]
      \centering
     \subfloat[]{
         \includegraphics[width=0.45\textwidth,draft=false]
         {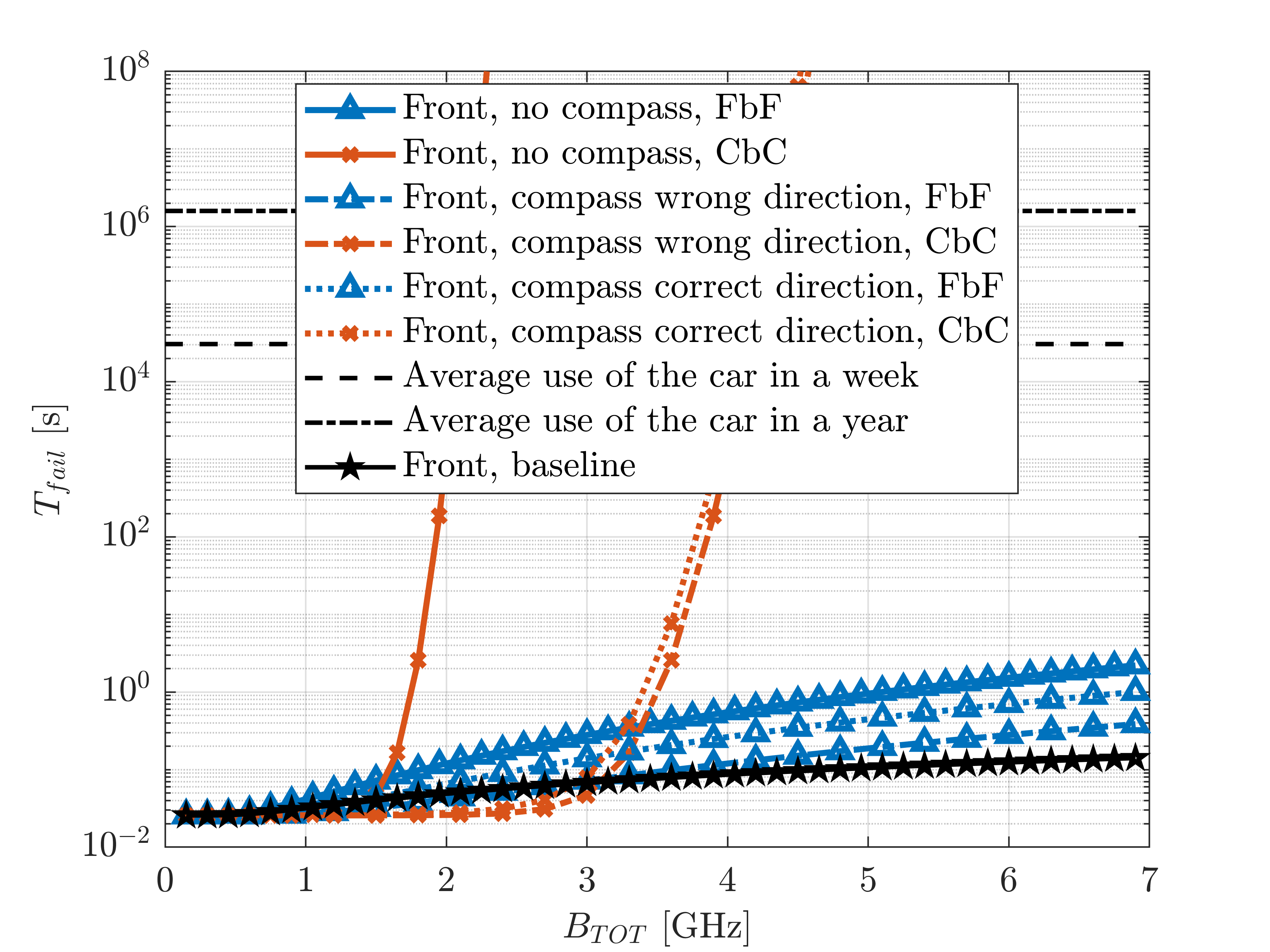}
          \label{fig:front_results_symmetric}
     }\subfloat[]{
     \includegraphics[width=0.45\textwidth,draft=false]{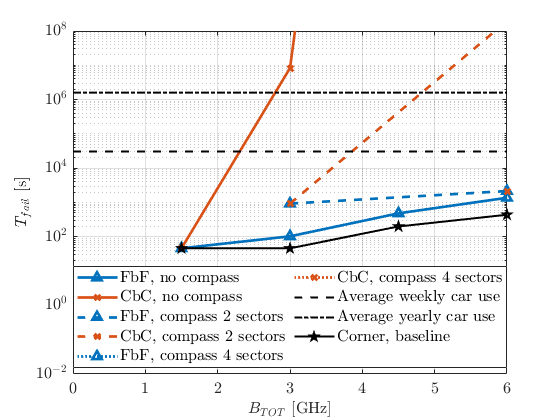}
          \label{fig:corner_results_symmetric}
    }\\[1mm]
        \caption{Average time between two system failures, varying the total available bandwidth, for different mitigation techniques in a scenario with 150 veh/km. Front (a) or corner (b) radars.}
        \label{fig:results_symmetric}
\end{figure*}

\begin{figure*}[t]
     \centering
     \subfloat[]{
         \includegraphics[width=0.45\textwidth,draft=false]{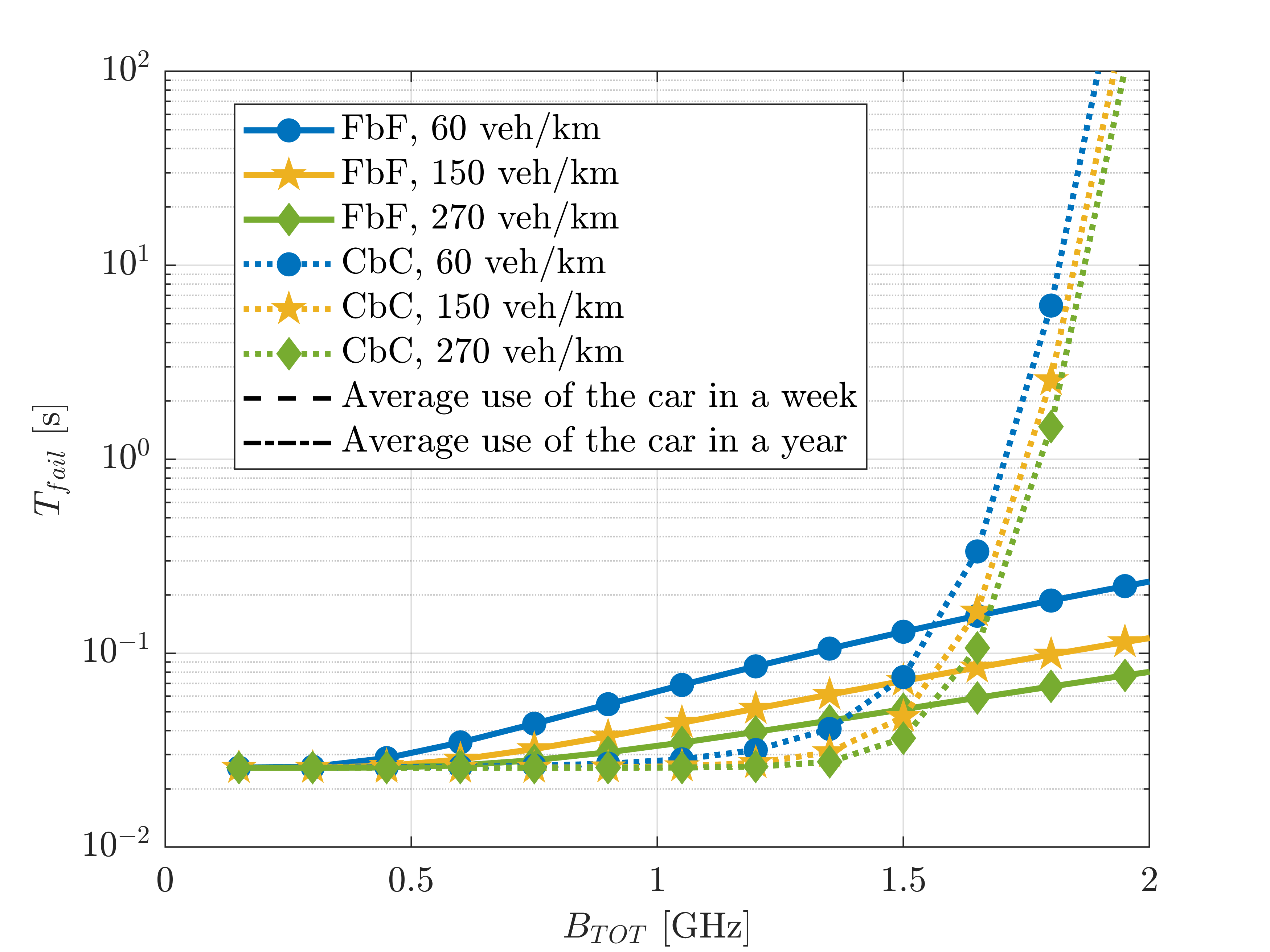}
          \label{fig:front_results_density_symmetric}
     }\subfloat[]{
         \includegraphics[width=0.45\textwidth,draft=false]{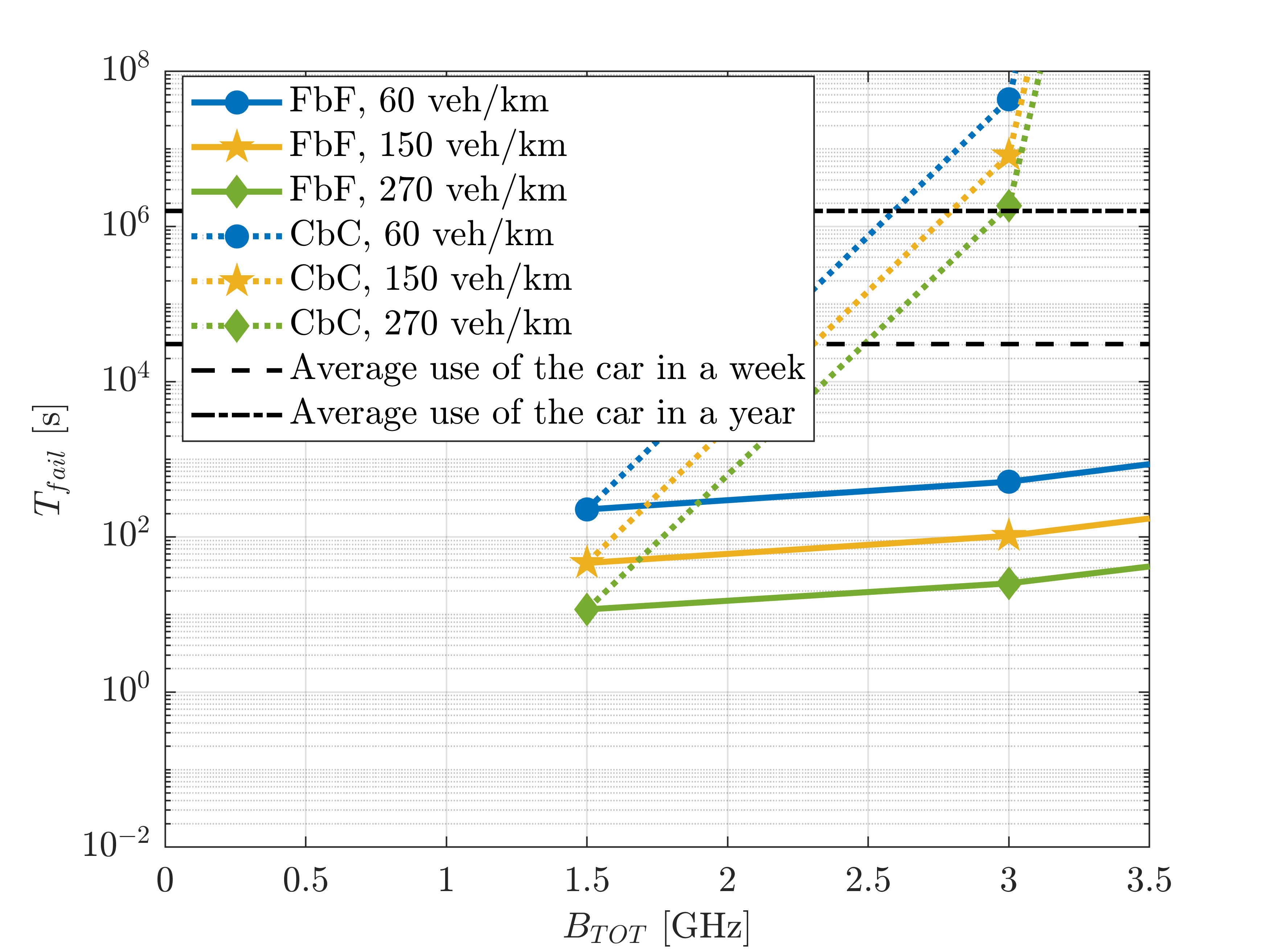}
          \label{fig:corner_results_density_symmetric}
     }\\[1mm]
        \caption{Average time between two system failures comparing chirp-by-chirp and frame-by-frame across multiple densities. Front (a) or corner (b) radars.}
        \label{fig:results_density_symmetric}
\end{figure*}

The impact of the various mitigation methods on the performance of the radar, using the settings in Table~\ref{tab:settings_1} is illustrated in Figs.~\ref{fig:front_PlotPe} 
to~\ref{fig:results_density_symmetric}. 

\rev{First, Fig.~\ref{fig:front_PlotPe} shows the probability that $M=3$ consecutive frames are lost as a function of the number $N$ of potential interferers and of the total available bandwidth. Specifically, Fig.~\ref{fig:front_PlotPe}(a) refers to the frame-by-frame case and the results correspond to $(\PcollF{N})^M$ (see \eqref{eq:PcollF}), while Fig.~\ref{fig:front_PlotPe}(b) refers to the chirp-by-chirp case, with values given by $(\PcollC{N})^M$ (see \eqref{eq:PcollC}). This figure enables inference of system behaviour for a fixed number of potential interferers, ranging from scenarios with few interferers, more likely for corner radars, to scenarios with several interferers, more likely for front radars. From Fig.~\ref{fig:front_PlotPe}, it can be observed that, as expected, the probability of $M$ consecutive frame losses increases with the number of potential interferers and decreases as the available bandwidth increases. Comparing frame-by-frame and chirp-by-chirp, note that the probability of $M$ consecutive losses under the chirp-by-chirp assumption is higher than that under the frame-by-frame assumption for a bandwidth of 1.5 GHz. However, this probability becomes negligible when the bandwidth is large, even in the presence of many potential interferers. Another important observation is that, in a pessimistic yet realistic scenario where the system bandwidth is limited to less than 1.5~GHz and more than 20 potential interferers are present, the probability of $M$ failures is well above 50\% for both mitigation methods.}

Then, in Fig.~\ref{fig:results_symmetric} the average time between failures $\Tfail$ is shown varying the available bandwidth $\BTOT$, assuming 150~veh/km and various mitigation methods. In particular, the blue curves correspond to the use of frame-by-frame frequency hopping with or without compass, the red curves to the use of chirp-by-chirp frequency hopping with or without compass, the black solid curve to the baseline, and the black non-solid curves to the reference average use of the car per week and year. In the case of front radars (Fig.~\ref{fig:front_results_symmetric}), the solid blue and red curves are used in the absence of compass, dashed curves if compass is used but the radars all fall in the same sector, and dotted curves if compass is used effectively. The second case mimics the worst situation where all radars share only half of the bandwidth, which may occur for example if the compass separates east from west and the cars all move on a south to north direction making them all allocated to either the east or the west sector. In the case of corner radars, instead, dashed curves are used for the use of two sectors and the dotted curves for the use of four sectors. As a general observation, it can be noted that all curves are equivalent when the bandwidth is very small and frequency hopping is not really possible.

Looking at the curves related to the front radars (Fig.~\ref{fig:front_results_symmetric}), if a small bandwidth is available, all solutions perform similarly, with the front radar without compass being the preferable solution. When the bandwidth is limited, in fact, randomly changing the frequency at every frame allows to have statistically some situations where the frame is not interfered (i.e., all potential interferers use a bandwidth that does not overlap by more than $\xf$). Hopping in frequency at every chirp in a small bandwidth increases the probability that at least $\kch$ chirps overlap in frequency and increases the probability of frame loss. After a certain total bandwidth, the chirp-by-chirp becomes very effective and the $\Tfail$ rapidly increases. 

By comparing the curves of Fig.~\ref{fig:front_results_symmetric} with and without compass, it is apparent that its application is counterproductive. The reduction of the number of potential interferers, observable in Figs.~\ref{fig:front_direct_equivalent_compass_symmetric} and~\ref{fig:front_bar_symmetric}, comes at the price that the bandwidth used in each sector is half that used without compass; the conclusion is that reducing the bandwidth has a negative effect which is more relevant than the gain due to the lower number of potential interferers.

\begin{figure*}[t]
      \centering
     \subfloat[]{
     \includegraphics[width=0.45\textwidth,draft=false]{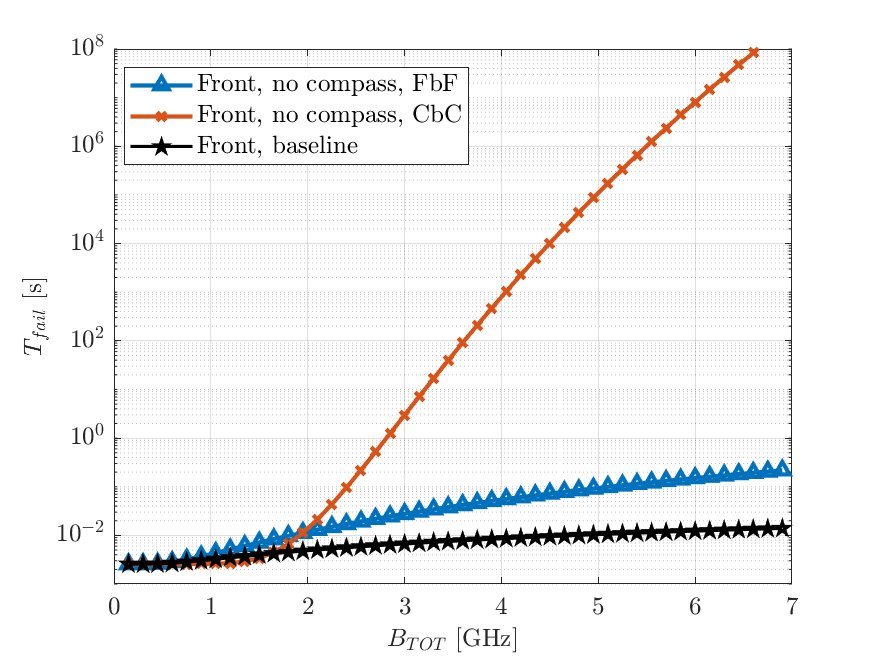}
          \label{fig:Additional200chirp}
    }
     \subfloat[]{
     \includegraphics[width=0.45\textwidth,draft=false]{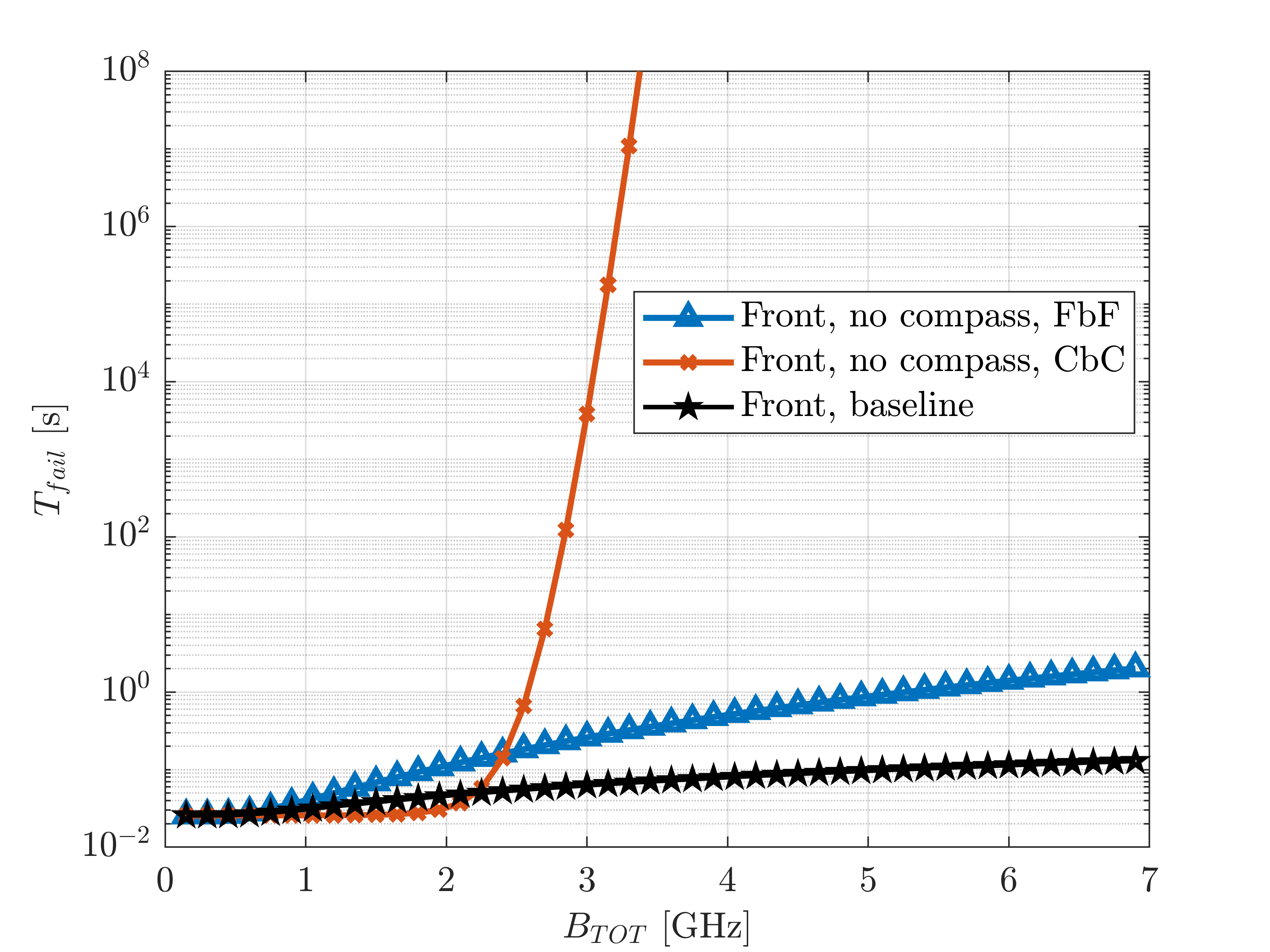}
          \label{fig:Additional150mhz}
    }
        \caption{\rev{Average time between two system failures with front radars, varying the  available bandwidth, for different mitigation techniques in a scenario with 150 veh/km, assuming different radar settings. (a) Assuming $N_\text{ch}=200$ and $K_\text{ch}=10$ (5\% of 200); (b) Assuming $B_\text{ADC}=150$~MHz}}
        \label{fig:additional_settings}
\end{figure*}

\begin{figure*}[t]
     \centering
     \subfloat[]{
         \includegraphics[width=0.45\textwidth,draft=false]{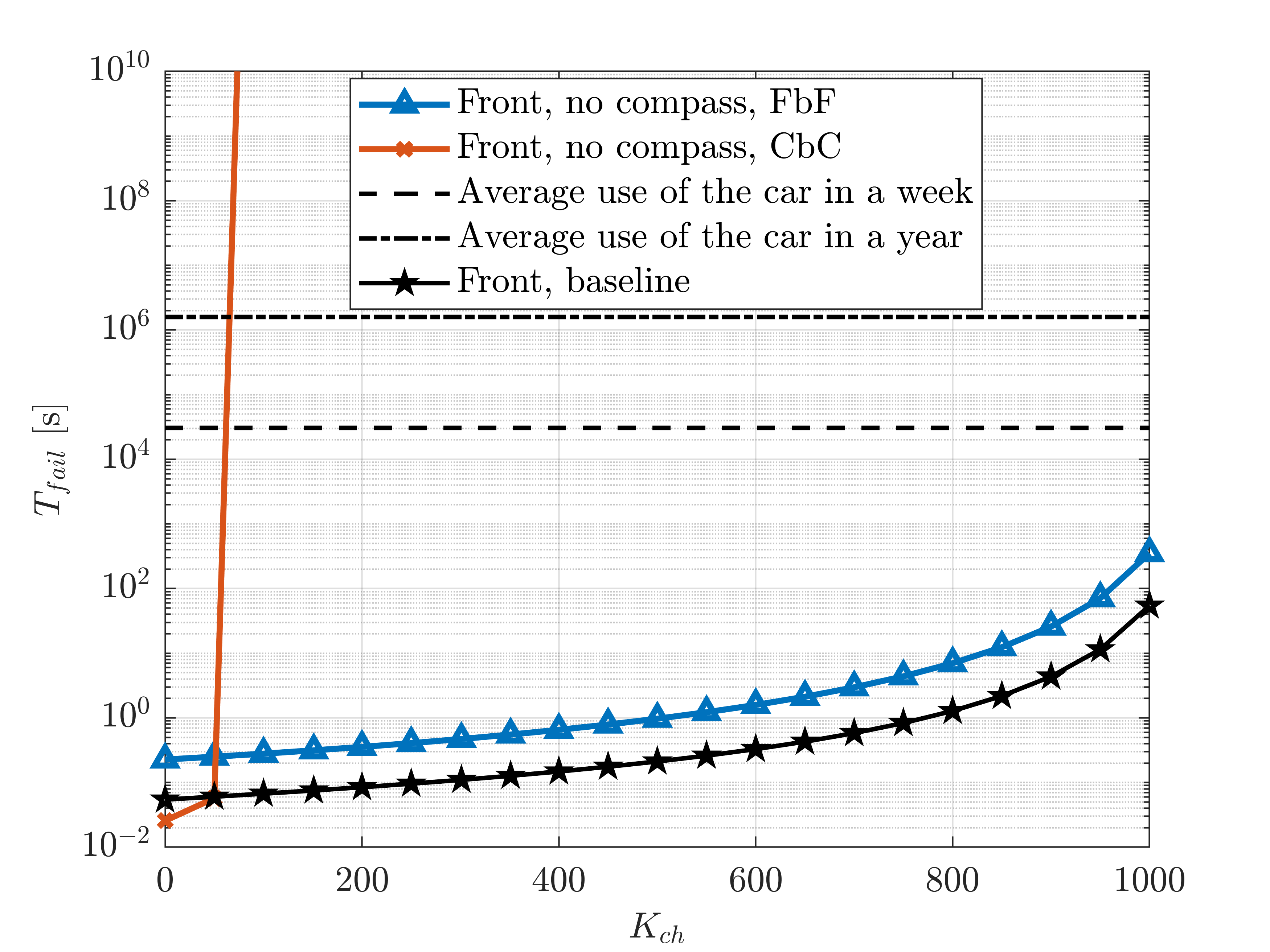}
          \label{fig:front_results_kch_symmetric}
     }\subfloat[]{
         \includegraphics[width=0.45\textwidth,draft=false]{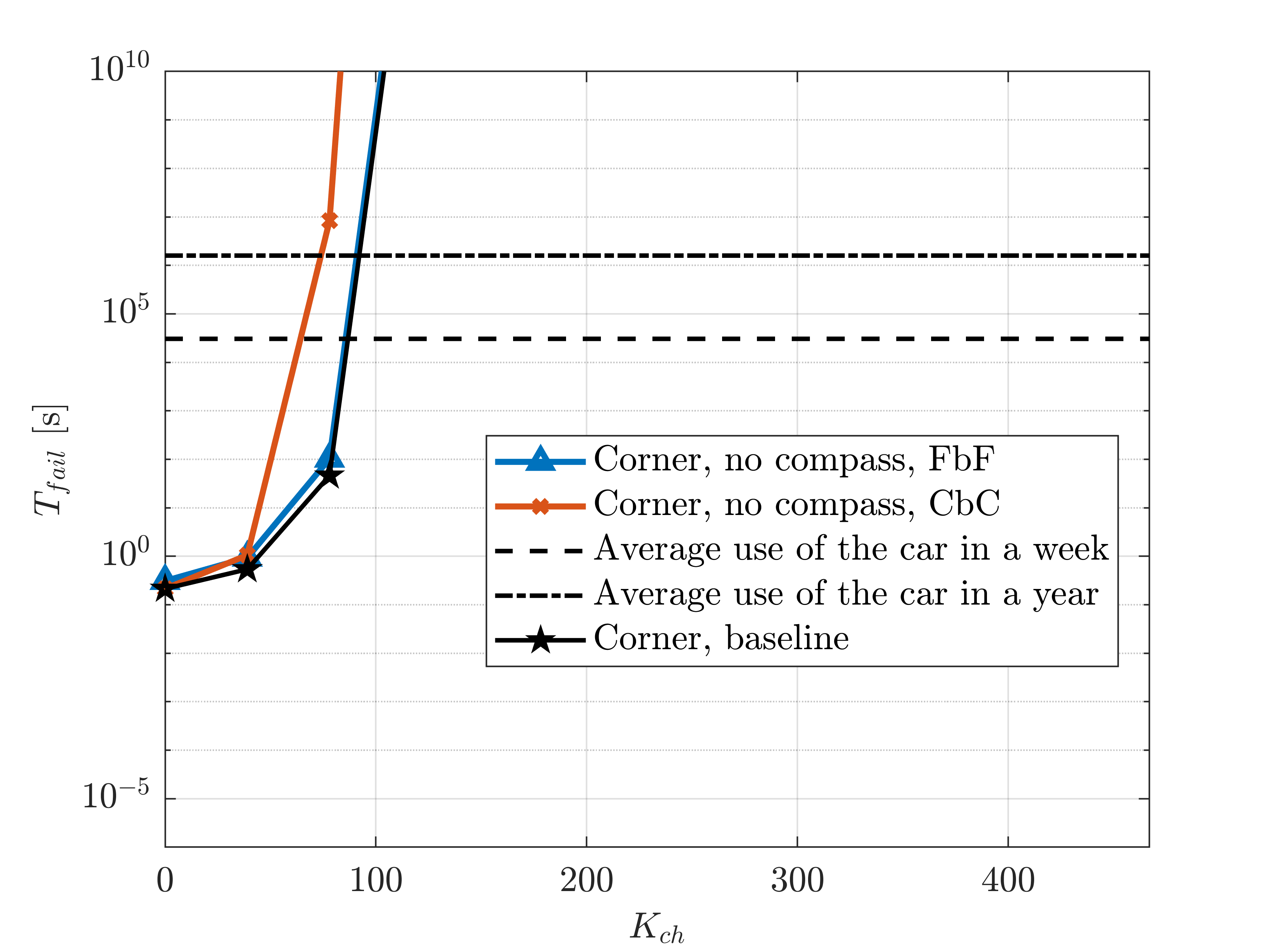}
          \label{fig:corner_results_kch_symmetric}
     }\\[1mm]
        \caption{Average time between two system failures, varying the number of chirps to be lost to have a frame loss, for different mitigation techniques in a scenario with 150 veh/km. Front (a) or corner (b) radars.}
        \label{fig:results_kch_symmetric}
\end{figure*}

\begin{figure*}[t]
      \centering
     \subfloat[]{
    \includegraphics[width=0.45\textwidth,draft=false]{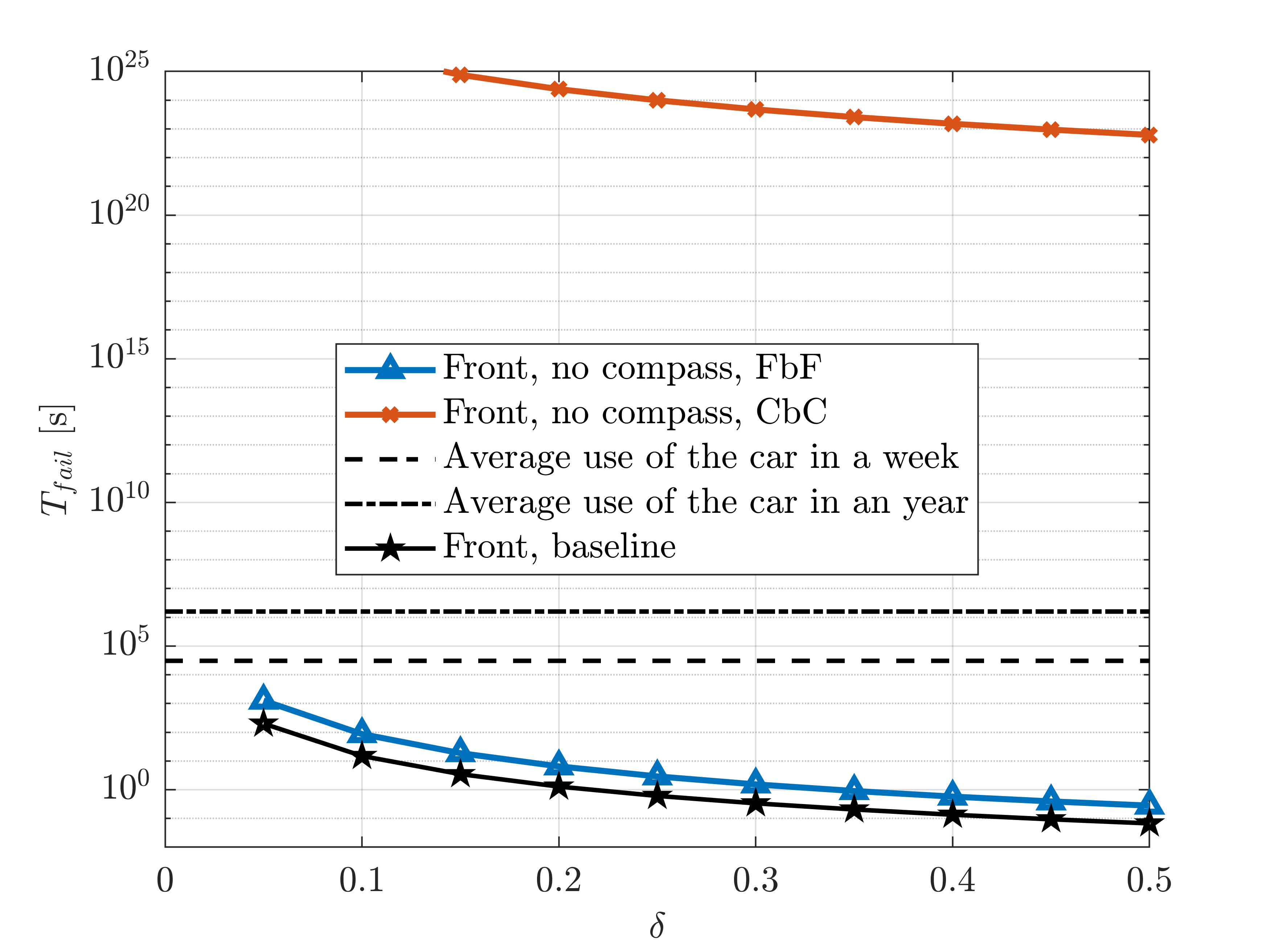}
          \label{fig:front_results_dutyCycle_symmetric}
     }\subfloat[]{
         \includegraphics[width=0.45\textwidth,draft=false]{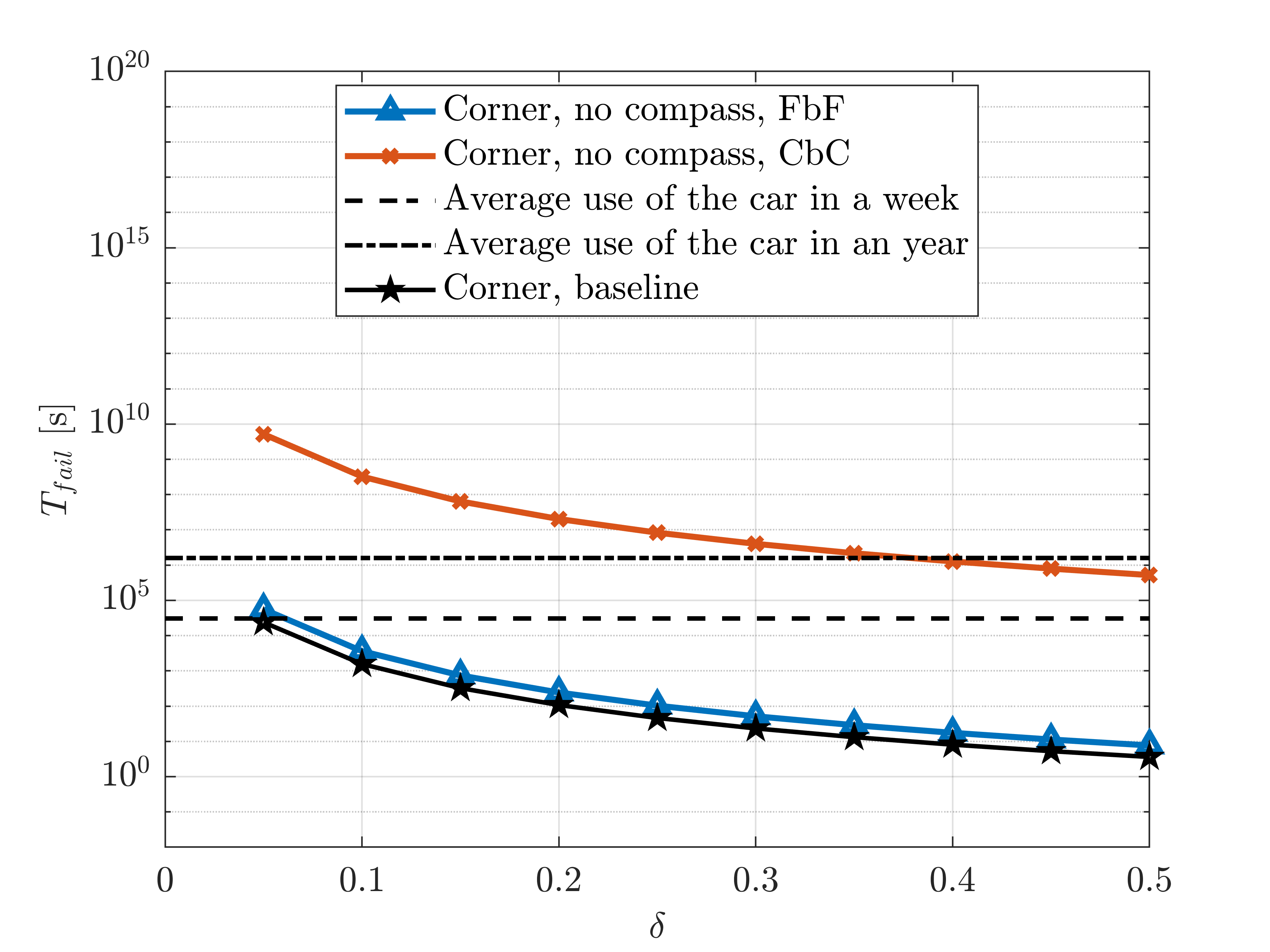}
          }\\[1mm]
        \caption{Average time between two system failures, varying the duty cycle, for different mitigation techniques in a scenario with 150 veh/km. Front (a) or corner (b) radars.} 
        \label{fig:results_dutyCycle_symmetric}
\end{figure*}

Moving to the corner radars (Fig.~\ref{fig:corner_results_symmetric}), results are similar. In this case, given the large chirp bandwidth, there are only few points in the plot. Related to this, it is relevant to notice that the use of compass with four sectors requires at least 6~GHz, thus giving results only in one point on the right of the figure; this configuration does not allow frequency hopping within the sector bandwidth. Still, frame-by-frame gives a slight increase of $\Tfail$ compared to the baseline, whereas the performance improvement allowed by chirp-by-chirp is significant. Compass is again counterproductive when added to chirp-by-chirp. In the case frame-by-frame, compass instead slightly helps, with two sectors preferable to four, consistently with the statistics of the potential interferers shown in Figs.~\ref{fig:corner_direct_equivalent_compass_symmetric} and~\ref{fig:corner_bar_symmetric}; the improvement appears however arguably sufficient when considering the increase of the system complexity.

Finally, Fig.~\ref{fig:results_density_symmetric} compares frame-by-frame and chirp-by-chirp without compass, under different traffic densities. These graphs are opportunely zoomed in to magnify the difference between curves in the semi-logarithmic domain. Overall, with a higher vehicle density the probability of system failure increases, but this risk appears well mitigated by employing the chirp-by-chirp technique with a sufficiently large bandwidth.

\subsection{\rev{Results changing the radar settings}}\label{sec:additional_settings}

\rev{The results shown in the previous subsection follows the settings detailed in Table~\ref{tab:settings_1}. To investigate the impact of some parameters and demonstrate that the observations made are of broader validity, in Figs.~\ref{fig:additional_settings} to~\ref{fig:results_dutyCycle_symmetric} additional results are shown assuming different radar settings.}

\rev{In particular, Fig.~\ref{fig:additional_settings} shows the performance varying the available bandwidth and assuming front radars, changing either the number of chirps or the ADC bandwidth. More specifically, Fig.~\ref{fig:Additional200chirp} follows the assumption that the number of chirps $N_\text{ch}$ is reduced to 200 (from 2000) and the number of chirps to cause a frame loss $K_\text{ch}$ is correspondingly reduced to 10 (from 100, remaining the 5\% of $N_\text{ch}$);
the shorter frame implies more frequent transmissions and for this reason the $\Tfail$ tends in general to reduce; furthermore, the chirp-by-chirp is slightly less effective due to the lower statistical variability that is introduced, especially when the system bandwidth is large. Fig.~\ref{fig:Additional150mhz} follows the assumption that the \ac{ADC} bandwidth $B_\text{ABC}$ is increased to 150~MHz (from 100~MHz); the larger ADC bandwidth slightly degrades the system performance. 
As expected, the variations introduced in the system settings make the absolute numbers to be different; however, the main conclusions remain the same: frame-by-frame is always preferable to baseline, and chirp-by-chirp allows for significantly better performance than the others, but only if the total available bandwidth is sufficiently large.}

Fig.~\ref{fig:results_kch_symmetric} then compares frame-by-frame and chirp-by-chirp, without compass, varying the parameter $\kch$ and fixing the total bandwidth to 3~GHz. As observable, this parameter significantly affects the performance of the system in terms of $\Tfail$. In particular, in the case of front radar, chirp-by-chirp almost completely cancels the risk of failures if the radar works properly with less than 10\% chirps corrupted (i.e., 200 of 2000); the performance of frame-by-frame instead improves but slower when $\kch$ increases. In the case of corner radar, the same tolerance of 10\% corrupted chirps (i.1., 156 of 1555 in this case) makes the interference negligible with all methods.

Again fixing the total bandwidth to 3~GHz, the impact of the duty cycle is finally analyzed in Fig.~\ref{fig:results_dutyCycle_symmetric}, where $\Tfail$ is shown versus $\duty$ assuming the mitigation methods without compass. As observable, for both front and corner radars, a reduction of the duty cycle allows to improve the performance and the impact is similar in all the cases. As an example, reducing $\duty$ from 0.5 to 0.25, the value of $\Tfail$ increases approximately by a factor 10.

\section{Conclusion}\label{Sec:conclusion}

The main conclusions deriving from the results can be summarized as follows:
\begin{enumerate}
    \item \textbf{Critical scenarios:} In terms of scenario, the most critical situations for front radars are those with a dense but not congested traffic; when the density is low, in fact, there are a limited number of interference sources; when the density is large, instead, the majority of sensors do not cause direct interference, because they are hidden by some vehicles acting as obstacles. Focusing on corner radars, the number of potential interferers, both direct and reflected, tends to increase as the traffic density rises. These conclusions can be inferred, for example, by looking at Fig.~\ref{fig:interferers_equivalent_symmetric}.
    \item \textbf{Impact of parameters:} As expected, the total available bandwidth is a key parameter for the study of the interference mitigation methods and can imply that one method is better than another, or the opposite (Figs.~\ref{fig:results_symmetric} and \ref{fig:results_density_symmetric}). The parameter $\kch$, which indicates the ability of the receiver to correctly detect the targets with a portion of the chirps corrupted by interference, is a critical parameter (Fig.~\ref{fig:results_kch_symmetric}); it should be noted that the value of $\kch$ used here (5\% of the chirps) is conservative and therefore the performance shown in this work appears pessimistic. The duty cycle also impacts on the performance, although in a less pronounced manner, and its reduction can help to improve the performance in critical situations (Fig.~\ref{fig:results_dutyCycle_symmetric}); 
    \item \textbf{Performance improvement with frequency-hopping:} Fig.~\ref{fig:results_symmetric} demonstrates the performance improvement achieved with the implementation of frequency-hopping techniques. Notably, all curves remain above the baseline, which represents the outcomes of radars using fixed frequencies (albeit randomly and independently set inside the overall available bandwidth);
    \item \textbf{Limitation of the compass method:} The compass method reduces the number of signals that can cause interference (see Figs.~\ref{fig:interferers_equivalent_compass_symmetric} and~\ref{fig:interferers_bar_symmetric}). However, this comes at the cost of reduced freedom in the bandwidth selected for the chirp transmission, which in turn limits the effectiveness of frequency variability to mitigate interference. As a result, in most cases, the use of compass appears counterproductive (see, in particular, Fig.~\ref{fig:results_symmetric}); 
    \item \textbf{Comparison of mitigation techniques:} 
    Using Fig.~\ref{fig:results_symmetric} as reference, the frame-by-frame technique outperforms the baseline under all conditions and appears as the best option when there is a limited available bandwidth; the chirp-by-chirp approach is particularly effective to mitigate the interference and appears overall the best option, 
    provided that (i) the number of potential interferers is not very large or (ii) the overall available bandwidth is sufficiently wide. 
\end{enumerate}


The overall message from this study is that frequency variations can significantly reduce the impact of interference, even in dense scenarios, but only if sufficient bandwidth is available. On the one hand, this may drive the investment needed to implement these methods in practice, and on the other hand, it 
justifies the current discussions on the need to reserve new bands for automotive radar. 
In future work, we plan to explore other mitigation methods, possibly including collision detection mechanisms, and to go beyond some of the assumptions made; in particular, we intend to consider different types of radar sharing the same bandwidth and the use of other signal waveforms instead of FMCW, such as \ac{OFDM}.

\input{Appendixes.tex}

\bibliographystyle{IEEEtran}
\bibliography{biblio}

\begin{IEEEbiography}[\vspace{-0.2cm}{\includegraphics[width=1in,height=1.2in,clip,keepaspectratio]{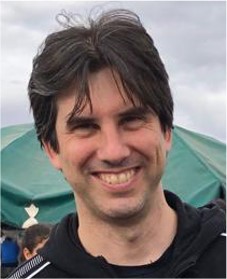}}]{Alessandro Bazzi}~(Senior Member, IEEE)  
is an Associate Professor at the University of Bologna. 
His research interests focus on networks of connected and autonomous vehicles, including radio resource management of wireless networks and large-scale automotive radar interference. On these topics, he coordinates projects and contributes to ETSI standardization.
\end{IEEEbiography}

\begin{IEEEbiography}[{\includegraphics[width=1in,height=1.2in,clip,keepaspectratio]{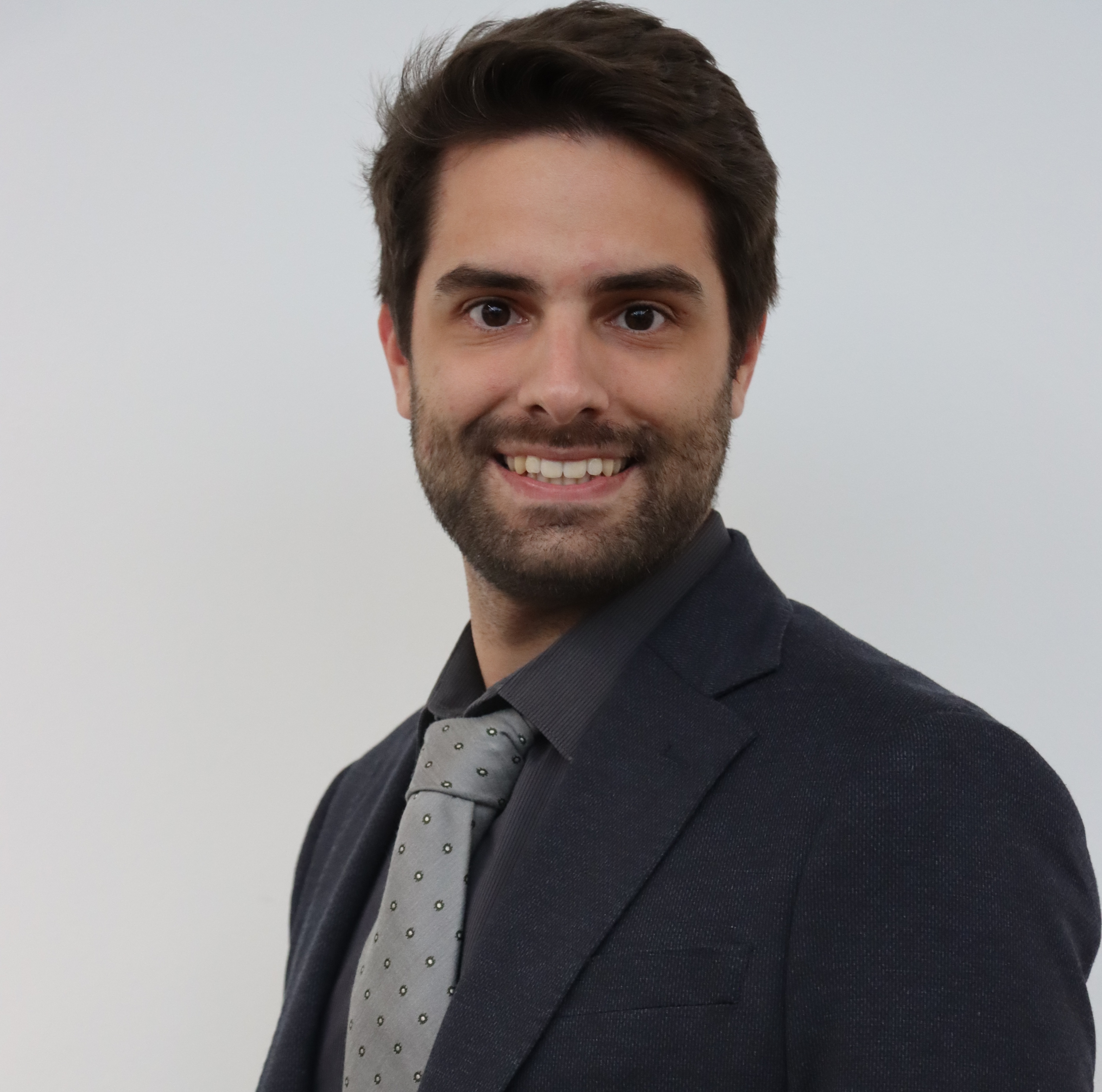}}]{Francesco Miccoli} received the B.S. degree in electronic engineering and the M.S. degree (summa cum laude) in electronic and telecommunications engineering from the University of Bologna, Italy, in 2020 and 2023, respectively. Since 2023, he has been a Senior Research Collaborator at the National Laboratory of Wireless Communications (WiLab) of CNIT (the National Inter-University Consortium for Telecommunications). His research interests include radar systems, advanced signal processing techniques, and wireless communications.
\end{IEEEbiography}

\begin{IEEEbiography}[{\includegraphics[width=1in,height=1.2in,clip,keepaspectratio]{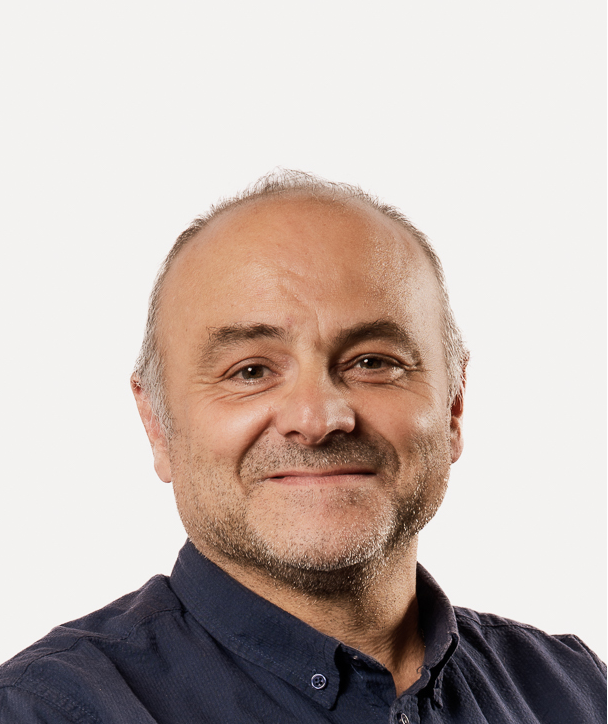}}]{Fabrizio Cuccoli}~(Senior Member, IEEE)  received the Laurea degree (cum laude) in electronic engineering in 1996 from the University of Florence and the Ph.D. in ”Methods and technologies for environmental monitoring” in 2001 from the University of Basilicata, Italy. From 2000 to 2009 he was researcher for the Interuniversity National Consortium for Telecommunications (CNIT) c/o the Department of Information Engineer ing, University of Florence, Italy. Since 2009 he has been Head of Research of the CNIT’s Radar and Surveillance systems (RaSS) Laboratory. His main research activity is in the area of remote sensing of rainfall, water vapor and atmospheric gaseous components through active systems and in the area of automotive radar.
\end{IEEEbiography}

\begin{IEEEbiography}[{\includegraphics[width=1in,height=1.2in,clip,keepaspectratio]{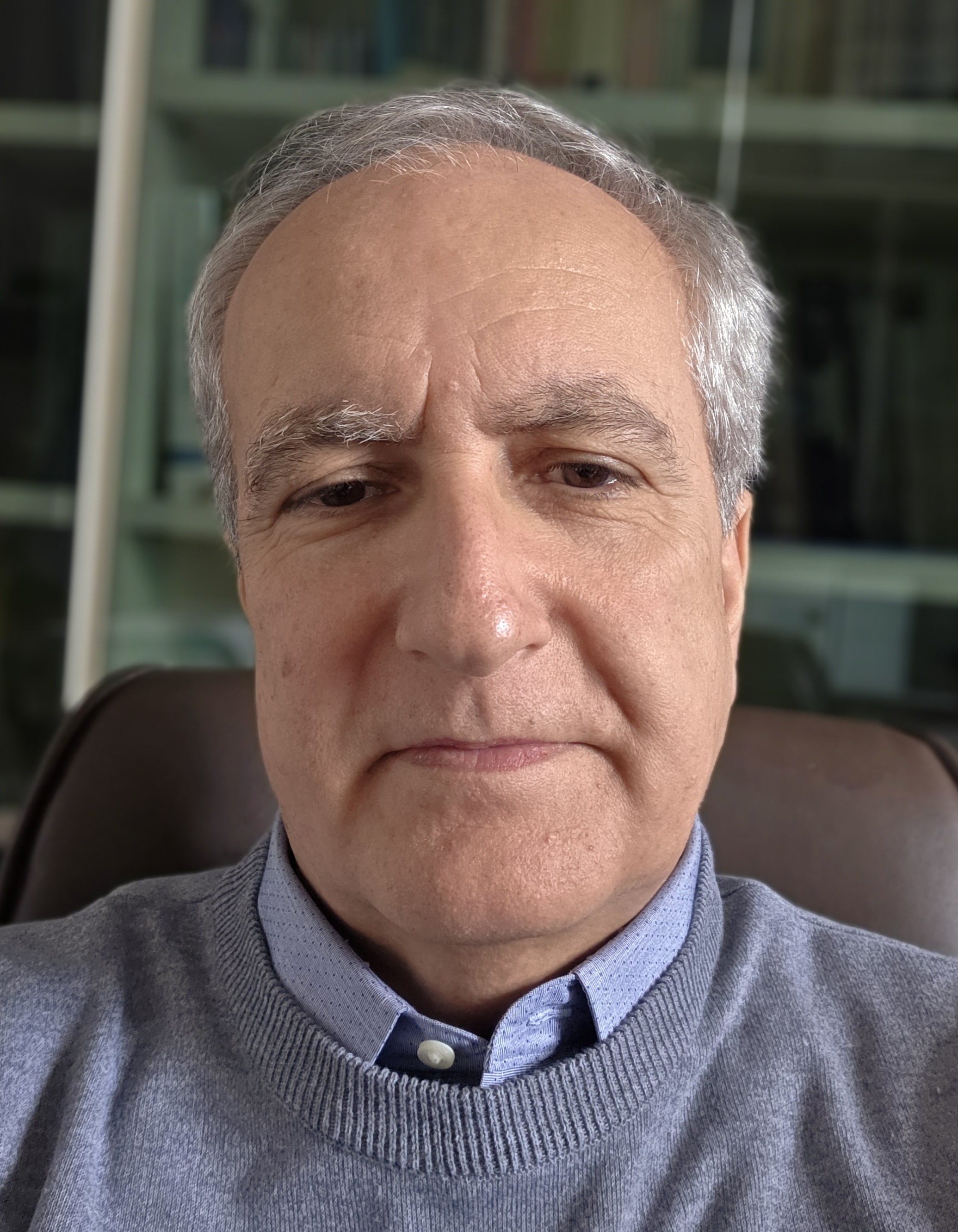}}]{Luca Facheris} received the Laurea degree (cum laude) in electronics engineering from the University of Florence, Italy, and the Ph.D. degree in electronics and information engineering in 1989 and 1993, respectively. From 1993 to 2002, he was an Assistant Professor with the Department of Information Engineering at the University of Florence, where he has been an Associate Professor since 2002. He teaches courses on radar systems, signal theory, information theory, and telecommunications. He has been involved in several research projects in his main areas of interest, which include radar systems and active atmospheric remote sensing.
\end{IEEEbiography}

\begin{IEEEbiography}[\vspace{-0.7cm}
{\includegraphics[width=1in,height=1.2in,clip,keepaspectratio]{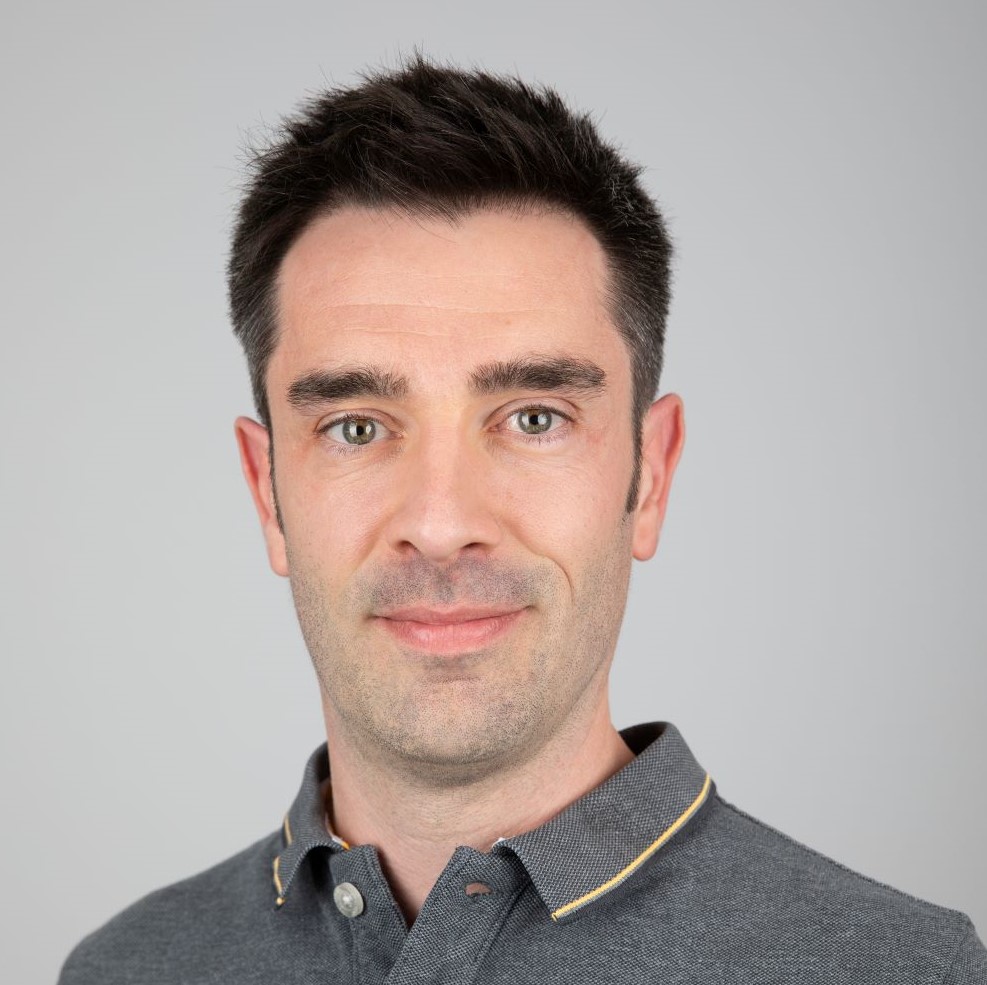}}]{Vincent Martinez} is a senior principal system engineer at NXP Semiconductors. He received a M.Sc. degree in mathematical modeling from INSA Toulouse, France. He is a wireless signal processing and physical layer expert, for automotive communication and radar applications. He is/has been contributing to various radar and ITS standardization and frequency regulation bodies.
\end{IEEEbiography}

\end{document}

%% file: acronyms.tex
\begin{acronym} 
\acro{ADAS}{advanced driver-assistance systems}
\acro{ADC}{analog to digital conversion}
\acro{AWGN}{additive white Gaussian noise}
\acro{CA-CFAR}{cell averaging constant false alarm rate}
\acro{CDF}{cumulative distribution function}
\acro{EIRP}{effective isotropic radiated power}
\acro{FOV}{field of view}
\acro{FMCW}{frequency modulated continuous waveform}
\acro{GNSS}{global navigation satellite system}
\acro{LOS}{line of sight}
\acro{NLOS}{non line of sight}
\acro{OFDM}{orthogonal frequency division multiplexing}
\acro{PPP}{Poisson point process}
\acro{RCS}{radar cross section}
\acro{SIR}{signal to interference ratio}
\acro{INR}{interference to noise ratio}
\acro{SINR}{signal to noise ratio}
\acro{SQL}{Structured Query Language}
\end{acronym}

%% file: symbols.tex
\newcommand{\Pfo}{{p_\text{f}}}
\newcommand{\xf}{{x_\text{f}}}
\newcommand{\BW}{{B_\text{ch}}}

\newcommand{\BTOT}{{B_\text{TOT}}}

\newcommand{\duty}{{\delta}}

\newcommand{\Ptchirp}{{p_\text{t-chirp}}}

\newcommand{\xt}{{x_\text{t}}}

\newcommand{\PtB}{{p_\text{t-frame}}}

\newcommand{\PcollB}[1]{{p_{\text{e(B)}_{|#1}}}}
\newcommand{\PcollF}[1]{{p_{\text{e(F)}_{|#1}}}}
\newcommand{\PcollC}[1]{{p_{\text{e(C)}_{|#1}}}}
\newcommand{\PfailB}{{p_{\text{fail(B)}}}}
\newcommand{\PfailF}{{p_{\text{fail(F)}}}}
\newcommand{\PfailC}{{p_{\text{fail(C)}}}}
\newcommand{\PfailX}{{p_{\text{fail}}}}

\newcommand{\Tchirprep}{{T_\text{rch}}}
\newcommand{\Tchirp}{{T_\text{ch}}}
\newcommand{\Mf}{{M}}

\newcommand{\xc}{{x_\text{t(C)}}}
\newcommand{\BADC}{{B_\text{ADC}}}

\newcommand{\kch}{{K_\text{ch}}}
\newcommand{\nch}{{N_\text{ch}}}

\newcommand{\Tfail}{{T_\text{fail}}}
\newcommand{\Tframe}{{T_\text{rf}}}
\newcommand{\PNint}[1]{{p_{\dmax,#1}}}
\newcommand{\PNintVector}{\mathbb{P}_{\dmax}}
\newcommand{\PNintRev}[1]{{p^*_{\dmax,#1}}}
\newcommand{\PNintRevVector}{\mathbb{P}^*_{\dmax}}

\newcommand{\varD}{D}
\newcommand{\varA}{A}
\newcommand{\varX}{x}
\newcommand{\pcoll}{p_\text{coll}}

\newcommand{\dmax}{\underline{d}}
\newcommand{\dref}{d_\text{ref}}
\newcommand{\RCS}{\sigma}

\newcommand{\pjzAppD}{{p_{j}}_{|z}}
\newcommand{\pframezAppD}{{p_{\text{frame}}}_{|z}}

\newcommand{\INRmindB}{{{\INRmin}}_\text{dB}}
\newcommand{\INRmin}{\gamma_\text{min}}
\newcommand{\light}{c}
\newcommand{\fzero}{f_0}
\newcommand{\EIRPdBm}{\epsilon_\text{dBm}}
\newcommand{\Gr}{{G_\text{r}}_\text{dB}}
\newcommand{\Boltzmann}{k_\text{B}}
\newcommand{\Tzero}{T_0}
\newcommand{\NF}{{F_\text{N}}_\text{dB}}

%% file: RelatedWorkSection.tex
\section{Related work}\label{sec:relatedwork}


The issue of interference in radar systems has been thoroughly investigated over the past decade, particularly in the automotive sector. A large number of studies have considered various waveforms (e.g., \cite{8768181,8792451}), and when focusing on \ac{FMCW}, they address both correlated and uncorrelated interference. However, in most cases, the studies limit their attention to a single interferer.
%
\input{TableRelatedWork} 
%
%
General considerations about the interference in FMCW radars are provided for example in \cite{4106078}, which presents real measurement examples, and in \cite{9266449},  where the topic is addressed through simulations. Additionally, \cite{8828037} and \cite{https://doi.org/10.1049/joe.2019.0167} propose parametric models that capture the effect of interference under various configurations. In all these cases, the authors assume a victim radar and a single interferer, noting that it is generally challenging to eliminate interference solely through signal processing at the receiver. They also highlight that interference, particularly when correlated, can lead to the generation of what they call 'ghosts' (i.e., false targets). 


Some work has also been done to investigate the impact of interference in scenarios with multiple vehicles. 
In \cite{7819520}, for example, the authors assume a scenario with a single lane in each direction and derive the statistics of the received interference, assuming either a 1-D \ac{PPP} or a lattice distribution. The performance of radars is then evaluated by comparing the \ac{SINR} with a certain threshold. In \cite{9745306}, the authors also calculate the distribution of the \ac{SINR}, assuming a 1-D PPP distribution and including Rayleigh fading. Based on this, they evaluate the delay in detection.  
In \cite{8904742}, 
the authors investigate the issue of interference through simulations, considering a two-lane highway with traffic moving in opposite directions, as well as a multi-lane highway with vehicles traveling in both the same and opposite directions. They also make general observations about the blockage effect and received interference. However, all these studies focus solely on direct interference, assume full signal overlap, and do not account for the specific waveform.


In addition to the studies that evaluate the interference,  
the literature also includes a large number of proposals for mitigating it, also developed during international projects such as MOSARIM and IMIKO \cite{9127843}. 
Focusing on proactive strategies, an overview of the main classes of solutions proposed to mitigate interference in FMCW-based radars is summarized in Table~\ref{tab:relatedwork}. A first group of solutions assumes the possibility of defining a communication channel where coordination is performed to reduce the risk of reciprocal interference. This can be achieved either through centralized coordination \cite{7733011,9070137,9266425} or direct exchanges between vehicles \cite{8835744,8943325,roudiere2021first,roudiere2021importance,10044192,9023460}. While these methods could potentially reduce interference significantly, the need for a communication channel and protocols to set parameters makes them difficult to implement in the short term. 
Similarly, using a carrier-sense type of channel access, as proposed in 
\cite{https://doi.org/10.1049/joe.2019.0166,9079146,k2022slotted}, also faces challenges. Also in this case, in fact, a common protocol would need to be standardized to avoid disrupting fair access to the medium. 
Given that these proposals are still in their early stages and require further investigation, an agreement cannot be expected for several years. A secondary, but important, aspect is that listen-before-talk methods are always studied under the assumption that all radars have the same settings. However, the likely scenario where radars have different chirp bandwidths or durations has not been explored.

There are also several solutions that do not appear easily applicable, as they alter the shape or power of the chirps, potentially affecting the predictability of radar performance. In particular, the proposals in \cite{s18092831,6544299,8768092} modify the shape of the chirps or introduce chirps with different characteristics, making it challenging to engineer the radar settings. Power control is instead proposed in \cite{8967012,9348730}; however, this approach is questionable as it may impact the radar range considered during design. Additionally, its effectiveness is debatable because if all radars reduce transmission power uniformly, the simultaneous reduction in both useful and interference power prevents any increase in the \ac{SINR}.

\rev{As a solution that would not alter radar performance and would not require particular coordination, also polarization had been proposed as an additional degree of orthogonality. In particular, the use of a +45 degree linear polarization was proposed in the MOSARIM project \cite{9127843} for front radars, with the idea that opposite radars would generate orthogonal signals. This approach, however, has limited effectiveness for several practical reasons. Among these, one is that radars are normally mounted behind bumpers, and this tends to cause polarization rotation; another is that the polarization has reduced impact if the source is lateral, which may cause the difference in polarization to be different from 90° and the orthogonality to be destroyed; additionally, using the same antennas for front, rear, and corner radars, is preferable, as designing and implementing different radars would be more expensive and logistically more challenging.}

In this work, we consider and compare the remaining approaches, which focus on varying the starting instant or frequency of the chirps. These methods 
can be implemented 
without altering radar performance during operation. The approaches examined are, in part, refinements of those proposed in \cite{9399786,9530166,5164283,10044244}. In particular, \cite{9399786} explores the use of random intervals between frames and the selection of one frequency band from a few available options for each chirp at random. However, this study considers only direct interference and an uncommon scenario where interference is generated by the preceding and following cars in the same lane as the target vehicle.
In \cite{9530166}, 
the authors propose dividing the time axis into slots to ensure that chirps transmitted in different slots do not interfere. They  
then assume that a random number of slots is waited after each chirp transmission before the next one.  The authors use ALOHA-type calculations to derive the probability of collision and simplify the scenario by assuming a fixed number of interferers. In \cite{5164283}, 
the authors study what they refer to as pseudo-noise modulation, 
which involves varying the starting frequency from chirp to chirp. The work appears to be preliminary and investigates performance through simulations, assuming a single source of interference. Finally, in \cite{10044244}, 
the overall bandwidth is divided into subchannels, with each chirp transmitted randomly in one of them. 
The results are obtained using a mathematical model applied to a multi-lane highway scenario. The authors calculate the average interference received in each lane and the probability of chirp overlap. These calculations are then used to compare the \ac{SINR} to a threshold and assess the accuracy of target detection.

In addition to the techniques that randomly select the starting instants and frequencies, we also consider the so-called compass method, first proposed within MOSARIM \cite{MOSARIM}. This method selects the starting frequency based on the radar's direction, which can be determined using \acp{GNSS}, for example. This technique has gained significant interest in recent years, as it promises to eliminate direct interference from radars pointed in directions different from that of the victim radar \cite{10149687}.

Unlike previous related work, we investigate here the effectiveness of various mitigation approaches as a function of the available bandwidth, assuming a realistic highway scenario and considering both direct and reflected interference from correlated sources. The evaluation uses mobility simulations to set the positions of the radars and then relies on mathematical models to calculate the probability of system failures.

It is worth noting that, although the considered methods offer clear theoretical advantages, they may be difficult to implement in practice. In particular, changing the chirp frequency arbitrarily or determining the vehicle’s direction to apply the compass method requires a not negligible implementation effort. For this reason, assessing the effectiveness of such methodologies when deployed on a large scale is of crucial importance.

%% file: TableRelatedWork.tex
\begin{table*}[]
    \centering
    \footnotesize
    \caption{Main strategies to mitigate the interference}\begin{tabular}{p{3cm}p{3cm}p{4.5cm}p{4.5cm}}
\hline
\hline 
\textbf{Strategy} & \textbf{References} & \textbf{Description} & \textbf{Considered}   \\
\hline  \hline
Controlled coordination & \mbox{\cite{7733011} (Khoury, 2016)} 
\mbox{\cite{9070137} (Huang, 2019)}
\mbox{\cite{9266425} (Mazher, 2020)} & A centralized entity sets the parameters based on the positions of the radars & NO, since it requires connectivity to the infrastructure and a controlling entity \\  \hline
Distributed coordination & \mbox{\cite{8835744,9127843,8943325}}(Aydogdu, 2019-21) \mbox{\cite{roudiere2021first,roudiere2021importance} (Roudiere, 2021)} \mbox{\cite{10044192} (Wang, 2023)} \mbox{\cite{9023460} (Zhang, 2020)} & Use of a separate communication channel to coordinate the settings of radar parameters (new or ITS-G5 or C-V2X sidelink) & NO, since it requires a common communication channel \\  \hline
Carrier sensing and backoff & \mbox{\cite{https://doi.org/10.1049/joe.2019.0166} (Kurosawa, 2019)} \mbox{\cite{9079146} (Ishikawa, 2019)} 
& The device senses if another radar is transmitting in the same band & NO, since it requires the definition of common rules \\  \hline
Sequences of chirp slopes & \mbox{\cite{s18092831} (Son, 2018)} & Use of sequences of chirp slopes that have low probability of mutual interference & NO, since it implies variable performance, needs coordination, and may require synchronization\\  \hline
Chirp-shape variability & \mbox{\cite{6544299} (Luo, 2013)} \mbox{\cite{8768092} (Umehira, 2019)} & Each chirp has different parameters, such as bandwidth or time duration & NO, since it implies different performance from one chirp to another \\  \hline
Power control & \mbox{\cite{8967012} (Chu, 2020)} \mbox{\cite{9348730} (Jin, 2020)} & The radar minimizes the used power to reach the required quality & NO, since it also reduces useful power, making it of questionable effectiveness\\   \hline
\rev{Polarization} & \rev{\mbox{\cite{MOSARIM} (MOSARIM, 2012)}} & \rev{Use of 45° linear polarization to make opposite directions orthogonal} & \rev{NO, since its application and effectiveness are limited in practice} \\ \hline
Variable starting frequency and/or time & \mbox{\cite{9399786} (Jin, 2021)} \mbox{\cite{9530166} (Haritha, 2021)} \mbox{\cite{5164283} (Mu, 2009)} \mbox{\cite{10044244} (Wang, 2023)} & Chirps start at different random frequencies or with variable starting time & YES, both as frame-by-frame and chirp-by-chirp 
\\  \hline
Compass & \mbox{\cite{MOSARIM} (MOSARIM, 2012)} \mbox{\cite{10149687} (Tovar Torres, 2023)} & The starting frequency is set based on the direction of the radar & YES \\  \hline
\hline
\end{tabular}

    \label{tab:relatedwork}
\end{table*}

%% file: TableSettings_1.tex
{
\begin{table}[t]
\caption{Radar system settings used inside the analytical models.  (*) indicates values used when not differently specified.}
\footnotesize 
\centering
\begin{tabular}{p{2.8cm}p{1.8cm}p{1.3cm}p{1.3cm}}\hline\hline
\textbf{Parameter}  & \textbf{Symbol} & \textbf{Front} & \textbf{Corner} \\ 
\hline
\hline
Duty cycle & $\duty$ & 0.5 (*)  & 0.25 (*)  \\
\hline
Duration of a chirp & $\Tchirp$ & 5.14 $\mu$s & 10.3 $\mu$s\\
\hline
Chirp repetition time& $T_\text{rch}$ & 6.42 $\mu$s & 12.8 $\mu$s\\
\hline
Frame repetition time 
& $\Tframe$ & 25.6~ms & 80~ms\\
\hline
Duration of a frame & $T_\text{f}$ & 12.8~ms & 20~ms \\
\hline
\# of chirps in a frame & $\nch$ & 2000 & 1555 \\
\hline
Available bandwidth & $\BTOT$ & 3 GHz (*) & 3 GHz (*) \\
\hline
Chirp bandwidth & $B_\text{ch}$ & 150 MHz & 1.5~GHz \\
\hline
Max. beat frequency  & $\hat{f}_\text{b}$ & 68.1 MHz & 97.29~MHz \\
\hline
ADC bandwidth & $\BADC$ & 100 MHz & 100 MHz \\
\hline
ADC sample frequency & $2 \cdot \hat{f}_\text{b}$ & 136.2 MHz & 194.6 MHz \\
\hline
Unambiguous Range & $R_\text{max} = \frac{c}{4}\cdot \frac{2\hat{f}_\text{b}\cdot T_\text{ch}}{B_\text{ch}}$ & 350.03 m & 100.21 m \\
\hline
Range resolution & $\Delta R = \frac{c}{2B_\text{ch}}$ & 1.00 m & 0.10 m \\
\hline
Maximum velocity & $v_\text{max} = \frac{c}{4f_{0}T_\text{rch}}$ & 83.44 m/s & 41.85 m/s \\
\hline
Velocity resolution & $\Delta v = \frac{c}{2f_\text{0}T_\text{f}}$ & 0.0834 m/s & 0.0538 m/s \\
\hline
Min. freq. overlap & $\xf$ & 0.5 & 0.5 \\
\hline
Chirp losses causing frame loss & $\kch$ & 100 (*) \hbox{(5\% of $\nch$)} & 78 (*) \hbox{(5\% of $\nch$)} \\
\hline 
Consec. frame losses & $\Mf$ & 3 & 3 \\
\hline
\hline
\end{tabular}
\label{tab:settings_1}
\end{table}
}

%% file: TableSettings_2.tex
{
\begin{table}[t]
\caption{Radar system settings used to find the maximum equivalent distance.}
\footnotesize 
\centering
\begin{tabular}{p{2.8cm}p{1.8cm}p{1.3cm}p{1.3cm}}\hline\hline
\textbf{Parameter}  & \textbf{Symbol} & \textbf{Front} & \textbf{Corner} \\ 
\hline
\hline
EIRP & $\EIRPdBm$ & 35 dBm & 15 dBm \\
\hline
Rx antenna gain & $\Gr$ & 30 dBi & 23 dBi\\
\hline
Field of view & FoV & 30° & 60° \\
\hline
Radar cross section & $\sigma$ & 10~$\text{m}^2$ & 10~$\text{m}^2$ \\
\hline
Noise figure & $\NF$ & 15 dB & 15 dB\\
\hline
Carrier frequency & $\fzero$ & 140 GHz & 140 GHz\\
\hline
ADC bandwidth& $\BADC$ & 100 MHz & 100 MHz \\
\hline
Minimum INR & $\INRmin$ & 0 dB & 0 dB \\
\hline
Maximum path length & $\dmax$ & 2694.90 m & 120.38 m \\
\hline \hline


\end{tabular}
\label{tab:settings_2}
\end{table}
}

%% file: Appendixes.tex
\section*{Appendix A: Probability of overlap in the frequency domain}\label{Chapter:AppendixB}


This Appendix details the derivation of \eqref{eq:Pfo}. 


Let us indicate the starting frequency of the victim and that of the attacker as $x$ and $y$, respectively, indicating them relatively to the starting frequency of the total available bandwidth. Both $x$ and $y$ are random variables, uniformly distributed within $0$ and $\BTOT-\BW$. We want to derive the probability that the two chirps of bandwidth $\BW$ overlap by at least $\xf\BW$, and this corresponds to calculate the probability that $|x-y|$ is lower than $\left(1-\xf\right) \BW$.

Given two variables uniformly distributed between $0$ and $M$ (where $M=\BTOT-\BW$ in our case), the probability density function of their difference is
\begin{align}
\text{Prob}\{x-y=z\} = f_z(z) =
\begin{cases}
\frac{M - |z|}{M^2} & -M \leq z \leq M \\
0 & \text{otherwise}
\end{cases}
\;.
\end{align}

The obtained function is a triangular function, symmetric with the y-axis. Thus, the probability that the absolute value of the difference is lower than a value $\Delta$ (where $\Delta = \left(1-\xf\right) \BW$ in our case) can be calculated as 
\begin{align}\label{eq:appIntegralTriangle}
&\text{Prob}\{|x-y|<\Delta\} = 2 \int_0^{\Delta} f_z(z) d(z) \nonumber \\
&= \left(\frac{M}{M^2} + \frac{M-\Delta}{M^2} \right) \cdot \Delta \nonumber\\
&= \frac{\Delta \left(2M - \Delta\right)} {M^2} = \frac{2 \Delta}{M}\frac{M - \frac{1}{2}\Delta}{M}
\end{align}

By introducing in \eqref{eq:appIntegralTriangle} $M=\BTOT-\BW$ and $\Delta = \left(1-\xf\right) \BW$ we obtain \eqref{eq:Pfo}.

\section*{Appendix B: Potential interferers with baseline}\label{Chapter:AppendixProbVect}



This Appendix details the derivation of \eqref{eq:PNintRev}.

Let us start from the probability vector $\PNintVector = \{\PNint{0},\PNint{1},\PNint{2},...\}$ of having none, one, two, etc. potential interferers, as defined in Section~\ref{sec:prodDistribution}. Our goal is to calculate the probability of having none, one, two, etc. potential interferers that overlap in frequency by at least a fraction $\xf$, assuming that each station always uses the same band. 

The resulting vector is denoted as $\PNintRevVector = \{\PNintRev{0},\PNintRev{1},\PNintRev{2},...\}$.
Given that the probability of overlapping by at least a fraction $\xf$ is $\Pfo$, 
%
%
the probability of having no potential interferers that overlap by at least $\xf$ 
can be calculated as
\begin{align}
    \PNintRev{0} & =  \PNint{0} + \PNint{1} \binom{1}{0} \left(\Pfo\right)^0 \left(1-\Pfo\right)^1 \nonumber\\& + \PNint{2} \binom{2}{0} \left(\Pfo\right)^0 \left(1-\Pfo\right)^2 + ...
\end{align}
where the first term represents the contribution from the original vector with no potential interferers, the second term accounts for the cases where there is one potential interferer which however does not overlap by a fraction $\xf$ or larger, 
the third term includes the cases where there are two potential interferers, but none overlap by a fraction $\xf$ with the victim, 
and so on.

Similarly, the probability of having one potential interferer that overlaps by at least $\xf$ 
can be calculated as
\begin{align}
    \PNintRev{1} & =  \PNint{1} \binom{1}{1} \left(\Pfo\right)^1 \left(1-\Pfo\right)^0 \nonumber\\& + \PNint{2} \binom{2}{1} \left(\Pfo\right)^1 \left(1-\Pfo\right)^1 + ...
\end{align}
where 
the first term includes all the cases where there is one potential interferer and it overlaps by at least $\xf$, 
the second term includes the cases where there are two potential interferers, with one that overlaps by at least $\xf$ and the other that does not, 
and so on.

By generalizing this calculation, we obtain \eqref{eq:PNintRev}. 

\section*{Appendix C: Probability of overlap in 
the time domain}\label{Chapter:AppendixD}

This Appendix details the derivation of \eqref{eq:Ptb}.

Let us assume one attacker, with the starting instants of the chirp repetition intervals aligned with those of the victim. This assumption, which simplifies the discussion, does not reduce the applicability of the model to a more general case, since the time variability is captured by the random position of the chirp in \eqref{eq:Ptchirp}. Each time there is an overlap of a chirp transmission interval, there is a collision with probability $\Ptchirp=\frac{\Tchirp}{\Tchirprep} \cdot \frac{\BADC}{\BW}$ (see \eqref{eq:Ptchirp}). Thus, if the frame from the attacker overlaps with the frame from the victim by $z$ chirp repetition intervals, the probability that there are $j$ collisions is 
\begin{equation}\label{eq_appD1}
\pjzAppD = \binom{z}{j} \Ptchirp^j (1-\Ptchirp)^{z-j}\;.
\end{equation}
Using \eqref{eq_appD1}, with $z$ chirp repetition intervals overlapping, the frame is lost if there are $\kch$ or more chirps collided and the probability of loosing the frame in that case becomes  
\begin{equation}\label{eq_appD2}
\pframezAppD = 1-\sum_{j=0}^{\kch-1} \pjzAppD\;.
\end{equation}

Given the aligned starting point of the chirp repetition intervals and considering that the frame duration and periodicity are assumed the same for all the radars, the starting point of the frame from the attacker can be in one of $\nch/\duty$ possible instants relatively to the starting instant of the frame of the victim, with uniform probability distribution (also recall that we assume that $\Tframe$ is a multiple of $\Tchirprep$). This means that the attacker frame starts in the same instant of the victim frame with probability $1/\left(\nch/\duty\right)$, one chirp repetition interval later with probability $1/\left(\nch/\duty\right)$, etc.

Since the frame of the attacker can start before or after the frame of the victim, there are two $\nch/\duty$ cases when the two transmissions overlap by one chirp repetition interval, two cases when the two transmissions overlap by two chirp repetition intervals, and so on. Varying the number of possible overlapping time intervals and using \eqref{eq_appD2}, the probability of loosing the frame is 

\begin{align}
\PtB& = \sum_{z=\kch}^{\nch-1}\frac{2}{\nch/\duty}\cdot \pframezAppD + \frac{1}{\nch/\duty}\cdot {p_{\text{frame}}}_{|\nch} \nonumber \\ &\simeq \sum_{z=\kch}^{\nch}\frac{2}{\nch/\duty}\cdot \pframezAppD
\end{align}
which corresponds to \eqref{eq:Ptb}. Please note that when $z=\nch$ there is only one case of overlap and not two; this would require the factor related to $z=\nch$ to be moved outside the summation. Since this negligibly impacts on the overall result, we prefer to approximate the equation to make it more compact.